\documentclass[prc,nofootinbib,aps,final,amsmath,amssymb,10pt,superscriptaddress,floatfix,letterpaper,showpacs]{revtex4-1}
\pdfoutput=1
\usepackage{eurosym}
\usepackage{graphicx}
\usepackage{amsmath}
\usepackage{amsfonts}
\usepackage{amssymb}
\usepackage[bookmarksnumbered,colorlinks,bookmarks,breaklinks=true,linkbordercolor={1 1 1},linktocpage,citecolor=blue]{hyperref}
\usepackage{graphicx,color}
\usepackage{subfigure}
\usepackage{dcolumn}
\usepackage{bm}

\newcommand\T{\rule{0pt}{2.6ex}}
\newcommand\B{\rule[-1.2ex]{0pt}{0pt}}
\def\etal{et al.~}
\def\etals{et al.}
\def\ql{``}
\def\qr{''\hspace{0.5mm}}
\def\qrs{''}

\def\UT{$^{235}$U}
\def\UF{$^{238}$U}

\def\PU{$^{239}$Pu}

\begin{document}

\title{Nucleon scattering on actinides using a dispersive optical model with extended couplings}
\author{E.\thinspace Sh.\thinspace Soukhovitski\~{\i}}
\affiliation{Joint Institute for Energy and Nuclear Research, 220109, Minsk-Sosny, Belarus}
\author{R.\thinspace Capote}
\email[Corresponding author, electronic address:~]{\underline{r.capotenoy@iaea.org}}
\affiliation{NAPC--Nuclear Data Section, International Atomic Energy Agency, Vienna
A-1400, Austria}
\author{J.\thinspace M.\thinspace Quesada}
\affiliation{Departamento de F\'{\i}sica At\'{o}mica, Molecular y Nuclear, Universidad de
Sevilla, Ap.1065, E-41080 Sevilla, Spain}
\author{S.\thinspace Chiba}
\affiliation{Research Laboratory for Nuclear Reactors, Tokyo Institute of Technology
2-12-1-N1-9 Ookayama, Meguro-ku, Tokyo 152--8550, Japan}
\author{D.\thinspace S.\thinspace Martyanov}
\affiliation{Joint Institute for Energy and Nuclear Research, 220109, Minsk-Sosny, Belarus}
\vspace{+0.5cm}
\begin{abstract}
Tamura coupling model (\textit{Rev.Mod.Phys.\thinspace}\textbf{17}\thinspace(1965)\thinspace679) has been extended to consider the coupling of additional low-lying rotational bands to the ground state band. Rotational bands are built on vibrational bandheads (even-even targets) or single particle bandheads (odd-$A$ targets) including both axial and non-axial deformations. These additional excitations are introduced as a perturbation to the underlying axially-symmetric rigid rotor structure of the ground state rotational band. Coupling matrix elements of the generalized optical model are derived for extended multi-band transitions in even-even and odd-$A$ nuclei. Isospin symmetric formulation of the optical model is employed.

A coupled-channels optical model potential (OMP) containing a dispersive contribution is used to fit simultaneously all available optical experimental databases including neutron strength functions for nucleon scattering on $^{232}$Th, $^{233,235,238}$U and $^{239}$Pu nuclei. Quasi-elastic ($p$,$n$) scattering data on $^{232}$Th and $^{238}$U to the isobaric analogue states of the target nucleus are also used to constrain the isovector part of the optical potential.
Lane consistent OMP is derived for all actinides if corresponding multi-band coupling schemes are defined. For even-even (odd-$A$) actinides almost all low-lying collective levels below 1~MeV (0.5~MeV) of excitation energy are coupled. OMP parameters show a smooth energy dependence and energy independent geometry. A phenomenological optical model potential that couples multiple bands in odd-$A$ actinides is published for a first time.

Calculations using the derived OMP potential reproduce measured total cross-section differences between several actinide pairs within experimental uncertainty for incident neutron energies from 50~keV up to 150~MeV. The importance of extended coupling is studied. Multi-band coupling is stronger in even-even targets due to the collective nature of the coupling; the impact of extended coupling on predicted compound-nucleus formation cross section reaches 5\%  below 3~MeV of incident neutron energy. Excitation of multiple bands in odd-$A$ targets is weaker due to the single-particle nature of the coupling. Coupling of ground-state rotational band levels in odd-$A$ nuclei is sufficient for a good description of the compound-nucleus formation cross sections as long as the coupling is saturated (a minimum of 7 coupled levels are typically needed).
\end{abstract}

\date{\today }
\maketitle

\section{Introduction}

The cross sections for the interaction of neutrons with actinide nuclei are crucially important for design of various fission reactor systems. Tight target uncertainties on the capture and inelastic scattering data for major actinides were derived from advanced reactor sensitivity studies by Nuclear Energy Agency WPEC Subgroup-26 \cite{WPEC-SG26:08}. However, the present status of evaluated nuclear data files for inelastic scattering on major actinides is not satisfactory as discussed at the IAEA Technical Meeting on \ql Inelastic Scattering and Capture Cross-section Data of Major Actinides in the Fast Neutron Region\qr \cite{INDC0597} held in 2011. Significant differences in evaluated inelastic cross sections were observed in evaluated
nuclear data files from 200 keV up to a few MeV of incident neutron energy for major actinides \cite{INDC0597}. Such differences lead to large uncertainties in the inelastic scattering cross section ($n,n^{\prime }$), as well as in calculated multiple neutron emission cross sections ($n$,2$n$) and ($n$,3$n$). The improvement of evaluated scattering cross sections and neutron emission spectra and reduction of their uncertainties for neutron induced reactions on actinides is an important issue that should initiate new theoretical studies.

Many theoretical models are being developed to describe nucleon scattering on actinide nuclei. The optical model is one of the fundamental theoretical tools which provides the basis of nucleon-scattering data evaluations \cite{hodg63}. An accurate calculation of the nucleon scattering from a well-deformed actinide nucleus using the optical model must include the coupling to the low-lying collective states. A very successful computational method to account for the importance of the multistep processes is the coupled-channels (CC) method \cite{bu63} using Tamura's formalism \cite{Tamura:65}, which permits an exact solution of the CC equations. Many of deformed optical potentials suggested for actinides are based on Lagrange work at CEA, France \cite{la75}; but coupled-channels equations were a formidable challenge for the computing capabilities of 70s and 80s. Many approximations were suggested and used; it was customary to couple only a few levels (usually 3) of the ground state rotational band in those
calculations. In 2004 two of the authors studied the convergence of neutron cross sections on $^{238}$U as a function of the number of coupled target states in the ground state rotational band. It was shown that the common practice of calculating neutron cross sections with 3 coupled levels in $K=0^{+}$ ground-state bands is inadequate \cite{Soukhovitskii:04}. This result was confirmed and generalized in a new comprehensive study by Dietrich \etal\cite{Dietrich:12}, but none of these papers included additional vibrational bands in the coupling scheme of even-even nuclei, nor additional rotational bands in odd-$A$ nuclei.

Needed higher accuracy of data for fast reactors requires improving the description of scattering data at incident neutron energies from a few keV up to 5--6 MeV to cover the region with the maximum yield of fission neutrons. While the energies of excited states of the ground-state band of even-even actinides below 500 keV are well described by a rigid rotor model, above 500 keV several vibrational bands are observed that need to be considered. The situation for odd actinides is even more complex, as no pairing gap exists, therefore low-lying excited states are dominated by rotational bands built on single-particle (1QP) bandheads (e.g., particle-hole configurations with $K=1/2^{+}$, $7/2^{-}$ and $5/2^{+}$, and corresponding rotational bands, dominate the low-lying excitation spectra for $^{235}$U and $^{239}$Pu nuclei). Therefore, low-lying rotational bands, built on vibrational bandheads for even-even targets, and on single-particle bandheads for odd-$A$ targets, need to be taken into account
to describe neutron inelastic scattering on actinides. This fact has long been recognized for even-even nuclei; a vibrational-rotational description within the coupled-channels approach has been used to describe scattering data on even-even actinides by University of Lowell group \cite{Chan:1982a,Chan:1982b,Sheldon:1986}, and later used by Kawano \etal\cite{Kawano:94}, Minsk group \cite{porsuk96,Maslov:06} and Bruy\`{e}res-le-Ch\^{a}tel group (e.g., Refs.~\cite{Lopez:05,Romain:16}). Authors presented preliminary findings of this work for axial-symmetric even-even nuclei in Ref.~\cite{Quesada:2013}. However, authors are not aware of similar published work undertaken for odd actinides\footnote{P.~Romain has used extended coupling schemes for calculations of the $n$+$^{235}$U reaction, but no results have been published.}, for which typically only the ground-state rotational band had been considered in optical model studies\footnote{Preliminary findings for odd-nuclei were presented at ND2013~\cite{Quesada:2015}, however current results supersede older publications as further understanding both in derivation and results was achieved.}. Additionally, those works used non-dispersive potentials to describe scattering data, except the dispersive potential used to describe neutron scattering on $^{238}$U target~\cite{Lopez:05} and other actinides~\cite{Romain:16}.

The main purpose of this comprehensive contribution is to improve the description of neutron scattering on actinides at incident neutron energies below 6~MeV by extending a previously derived dispersive optical model potential for nucleon induced reactions on actinides based on a rigid-rotor description \cite{socaqu05,casoqu05,Capote:08}. That previously published by authors rigid-rotor potential of Refs.~\cite{socaqu05,casoqu05,Capote:08} will be referred in the rest of this work as the \ql RIPL 2408\qr potential by using the keyword from the RIPL optical-model database \cite{ripl3}.

Special focus will be on predicted compound-nucleus formation cross section $\sigma_{CN}(E)$, which is a critical input quantity for statistical reaction modelling. It has been shown that $\sigma_{CN}(E)$ strongly depends on employed coupled-channels couplings~\cite{Dietrich:12}. The proposed extension should account for multiple-band couplings, while keeping the achieved quality of description of the whole set of scattering data (e.g., total cross section data above 5~MeV were described by the RIPL~2408 potential within the quoted experimental uncertainty of about 1.5\%). The coupling of vibrational bands is expected to improve the description of neutron scattering on even-even actinides in the energy region from 500~keV up to about 6~MeV, which is critical for fast neutron fission reactors. The coupling of rotational bands built on single-particle bandheads for odd-$A$ nuclei is expected to reduce observed discrepancies from 100~KeV up to 1~MeV incident neutron energy in inelastic scattering on major fissile actinides $^{233}$U, $^{235}$U and $^{239}$Pu \cite{INDC0597}. The authors are unaware of previously published optical model studies for odd-$A$ actinide targets involving coupling of multiple bands.

The paper is structured as follows. Sections II and III provide descriptions of the nuclear shape parametrization, nuclear hamiltonian and nuclear wave functions needed in the solution of the Schr\"odinger equation describing the nucleon scattering problem. Section IV contains the new coupling matrix elements derived within Tamura's formalism allowing for multiple band coupling; this is one of key results of this paper. For interested readers a detailed derivation is given in Appendices 1 and 2.
Section V describes the formulation of dispersive Lane-consistent potential for the simultaneous description of neutron and proton induced reactions on actinides. In section VI, we discuss derived optical potential parameters to describe nucleon scattering on actinides using a proper multi-band level scheme; we also compare the derived potential with selected
experimental data, including highly accurate averaged total cross-section differences measured for incident neutron energies
from 50~keV up to 150~MeV. Comprehensive coupling schemes and multi-band coupling parameters derived from a least-square fit for $^{232}$Th, $^{233,235,238}$U, and $^{239}$Pu targets are tabulated in Appendix~3. Finally, section VII contains our conclusions.

\section{Nuclear shape parameterization for extended coupling scheme\label%
{nuclear_shape}}

Actinides are well deformed nuclei, where low-lying collective levels are strongly excited in nucleon inelastic scattering. The employed nuclear shape parameterization combines a rigid rotor description of the ground state rotational band, with small 
quadrupolar and octupolar vibrations around the deformed equilibrium shape. Vibrations are described in the spirit of the soft-rotator model \cite{SRM1,SRM2}, which considers nuclei to be deformable with both axial and non-axial vibrations; deformations with $\lambda \geqslant 4$ are considered axial, assuming that such deformations are usually small. Rotational bands are built on top of vibrational (single-particle) bandheads for even-even (odd) nuclei, respectively. The deformed nuclear optical potential arises from deformed instant nuclear shapes of actinide nucleus. The instant nuclear shape in a body-fixed system can be described as follows
\begin{eqnarray}
R_i(\theta ^{\prime },\varphi ^{\prime }) &=&R_{0i}^{{}}\Bigg\{1+\underset{%
\lambda =4,6,8}{\sum }\beta _{\lambda 0}Y_{\lambda 0}(\theta ^{\prime })
\notag \\
&&+\beta _{2}\left[ \cos \gamma Y_{20}(\theta ^{\prime })+\frac{1}{\sqrt{2}}%
\sin \gamma ~\left[ Y_{22}(\theta ^{\prime },\varphi ^{\prime
})+Y_{2-2}(\theta ^{\prime },\varphi ^{\prime })\right] \right] +
\label{SRM_R} \\
&&+\beta _{3}\left[ \cos \eta Y_{30}(\theta ^{\prime })+\frac{1}{\sqrt{2}}%
\sin \eta ~\left[ Y_{32}(\theta ^{\prime },\varphi ^{\prime
})+Y_{3-2}(\theta ^{\prime },\varphi ^{\prime })\right] \right] \Bigg\},
\notag
\end{eqnarray}%
where $Y_{\lambda \mu }(\theta ^{\prime },\varphi ^{\prime })$ means spherical harmonics; $\theta ^{\prime }$ and $\varphi ^{\prime }$ are the angular coordinates in the body-fixed (intrinsic) system; $\beta _{\lambda
0} $ are the static axial deformations, $\beta _{2}$ and $\beta _{3}$ are the
quadrupolar and octupolar deformations, and $\gamma $ and $\eta $
are the quadrupolar ($\gamma $) and octupolar ($\eta $) non-axial
deformation parameters, respectively. We introduced the notation $Y_{\lambda
0}(\theta ^{\prime })\equiv Y_{\lambda 0}(\theta ^{\prime },\varphi ^{\prime
})$. We followed Bohr \cite{Bohr52} and assumed that the body-fixed frame is
chosen to be the principal nuclear symmetry axes, such that $\beta _{21}=\beta _{2-1}=0$ and $%
\beta _{22}=\beta _{2-2}$, therefore quadrupolar deformations are described
by two parameters $\beta _{2}$ and $\gamma $. We also followed Lipas and
Davidson \cite{Lipas:1961} and assumed that octupolar vibrations with even
projection ($\lambda =3,\mu =0,\pm 2$) best describe the octupole
contribution to the nuclear vibrations for the low-lying negative parity
states. Therefore, the non-axial deformations $\beta _{3\pm
1}=\beta _{3\pm 3}=0$, and octupolar deformations are described by two
parameters $\beta _{3}$ and $\eta $ (such assumption could be easily removed
if needed by introducing one additional parameter).
In what follows we will use the fact that for nonaxial contributions ($\mu \neq 0$) in Eq.~\eqref{SRM_R} only even values of the projection $\mu =0,\pm 2$ are allowed both for quadrupolar $\lambda =2$ and octupolar $\lambda =3$ vibrations.

Nuclear radii $R_{0i}=r_{i}A^{1/3}$ ($i=HF,v,s,C,so$) are defined for five potential geometries described in the next section, being $A$ the target mass number. The soft rotator model of nuclear structure has been successfully applied in coupled-channels  optical model analyses for many nuclei \cite{porsuk96,porsuk98,Lee:2011}. However, until recently we were unable to derive a \textit{dispersive} coupled-channels optical model potential for actinides based on soft-rotator couplings. Such work is still in progress.
Dispersive optical model features a reduced number of parameters compared to traditional optical-model analyses, therefore parameter compensation of model defects become more difficult for dispersive potentials. Mathematically, the failure of the traditional soft-rotator model applied to actinides is due to the very slow convergence of the multipolar expansion of the soft-rotor potential around the spherical shape for large equilibrium deformations typically encountered in actinides. To solve this problem it was assumed that the quadrupole variable $\beta _{2}$ of the soft rotator model can be considered as a large equilibrium axial component $\beta _{20}$ plus a small contribution $\delta \beta _{2}$ (i.e. $\beta _{2}=\beta _{20}+\delta \beta_{2}$). This assumption allows a much better convergence of the potential expansion, as we use a truly small parameter $\delta \beta _{2}$, instead of the relatively large parameter $\beta _{2}$ originally used by Tamura \cite{Tamura:65}. The departures from the axially-symmetric rigid-rotor shape are considered to all orders in quadrupole non-axiality parameter $\gamma $ and octupole non-axiality parameter $\eta $. 
Under these assumptions we can rewrite the Eq.~\eqref{SRM_R} as follows
\begin{eqnarray}
R_{i}(\theta ^{\prime },\varphi ^{\prime }) &=&R_{0i}^{{}}\left\{ 1+\underset%
{\lambda =2,4,6,8}{\sum }\beta _{\lambda 0}Y_{\lambda 0}(\theta ^{\prime
})\right\} +R_{0i}^{{}}\beta _{20}^{{}}~\left[ \frac{\delta \beta _{2}}{%
\beta _{20}}\cos \gamma +\cos \gamma -1\right] Y_{20}(\theta ^{\prime })+
\notag \\
&&R_{0i}^{{}}(\beta _{20}^{{}}+\delta \beta _{2})\frac{\sin \gamma }{\sqrt{2}%
}\left[ Y_{22}(\theta ^{\prime },\varphi ^{\prime })+Y_{2-2}(\theta ^{\prime
},\varphi ^{\prime })\right] +R_{0i}^{{}}\beta _{3}\cos \eta Y_{30}(\theta
^{\prime })+  \label{radius} \\
&&R_{0i}^{{}}\beta _{3}\frac{\sin \eta }{\sqrt{2}}\left[ Y_{32}(\theta
^{\prime },\varphi ^{\prime })+Y_{3-2}(\theta ^{\prime },\varphi ^{\prime })%
\right].  \notag
\end{eqnarray}

The first term of the sum in curly brackets in Eq.~\eqref{radius} corresponds to the axially-symmetric equilibrium shape (rigid rotor). There are two additional quadrupole and octupole axially-symmetric terms (proportional to $Y_{20}(\theta ^{\prime })$ and $Y_{30}(\theta ^{\prime })$, respectively), and non-axial quadrupole and octupole terms. If we further assume that the octupole deformation parameter $\beta _{3}$ is small, we can expand the nuclear potential $V(r,R(\theta ^{\prime},\varphi ^{\prime }))$ around the equilibrium rigid rotor shape (instead of the equilibrium spherical shape used by Tamura -- see Eq.~(5) of Ref.~\cite{Tamura:65}). If we insert the nuclear shape given by Eq.~\eqref{radius} into the optical potential, and make a first order Taylor expansion around the equilibrium (axial) shape, we can obtain the following expression for the deformed optical
potential expansion%
\begin{eqnarray}
V(r,R(\theta ^{\prime },\varphi ^{\prime })) &=&\left[ V(r,R(\theta ^{\prime },
\varphi ^{\prime }))\right] _{(\delta \beta _{2}=0,\gamma =0,\beta _{3}=0)}+%
\left[ R_0 \frac{\partial }{\partial R}V(r,R(\theta ^{\prime },\varphi ^{\prime
}))\right] _{(\delta \beta _{2}=0,\gamma =0,\beta _{3}=0)}  \notag \\
&&\times \Bigg\{\beta _{20}\left[ \frac{\delta \beta _{2}}{\beta _{20}}\cos
\gamma +\cos \gamma -1\right] Y_{20}(\theta ^{\prime })+(\beta _{20}+\delta
\beta _{2})\frac{\sin \gamma }{\sqrt{2}}\left[ Y_{22}(\theta ^{\prime
},\varphi ^{\prime })+Y_{2-2}(\theta ^{\prime },\varphi ^{\prime })\right]
\label{pot_expansion} \\
&&+\beta _{3}\left[ \cos \eta Y_{30}(\theta ^{\prime })+\frac{\sin \eta }{%
\sqrt{2}}\left[ Y_{32}(\theta ^{\prime },\varphi ^{\prime })+Y_{3-2}(\theta
^{\prime },\varphi ^{\prime })\right] \right] \Bigg\}  \notag \\
&&+\text{ (potential higher derivatives' terms)},  \notag
\end{eqnarray}%
where
\begin{itemize}
\item
$\left[ V(r,R(\theta ^{\prime},\varphi ^{\prime }))\right] _{(\delta \beta
_{2}=0,\gamma =0,\beta _{3}=0)}\equiv V_{rot}(r,R_{axial}(\theta ^{\prime
})) $,
\item $\left[ R_0 \frac{\partial }{\partial R}V(r,R(\theta ^{\prime
},\varphi ^{\prime }))\right] _{(\delta \beta _{2}=0,\gamma =0,\beta
_{3}=0)}\equiv \left[ R_0 \frac{\partial }{\partial R}
V(r,R(\theta ^{\prime},\varphi ^{\prime }))\right] _{R=R_{axial}(\theta^{\prime })}$,
\end{itemize}
\noindent are the axially-symmetric components of the deformed optical model potential, and the non-axial components are included in Eq.~\eqref{pot_expansion} within curly brackets.
A general expression for the multipolar expansion of the deformed optical model potential is derived in Appendix~1, see Eq.~\eqref{V_multip_exp}.

\section{Nuclear Hamiltonian and target-nucleus wave functions}
Excited states in actinides are usually classified into single-particle and collective excitations, being the latter divided into vibrational and rotational states. Such division works well for even-even nuclei due to the pairing gap; only collective states are located inside the gap. However, for odd-$A$ actinides the energy of single-particle excitations could be of the same order of the rotational energies, and so the independence of the single-particle excitations from the collective movement is hardly justifiable. These cases have been extensively studied by Davidov and coworkers \cite{davidov60,davidov61,davidov62,davidov-book}, and Davidson \cite{davidson65} (for additional information see references therein). Davidov showed that in even-even nuclei whose equilibrium shape is close to the axially-symmetric shape (e.g., actinides), rotation cannot be regarded independently of $\gamma $-vibrations \cite{davidov61}. For odd nuclei, the interaction of the nuclear rotation with the unpaired nucleon may change both the structure of the rotational spectra and the energies of the single-particle excited states \cite{davidov-book}. Therefore, it is expected that low-lying excited states of odd-$A$ nuclei have a complex structure which does not permit the separation of the single-particle, rotational and vibrational degrees of freedom. However, all excited states in actinides are expected to have definite values of parity $\pi =(-1)^{l}$, being $l$ the orbital angular momentum, and of the total angular momentum $I$. The analysis above clearly showed that even-even and odd-$A$ actinides have a very different nuclear structure.

Let's define the nuclear structure model to be used in this work. We assume that a
nucleus is comprised of the even-even core where only paired nucleons are
present. Additionally, for odd-$A$ nuclei, we consider a single unpaired
nucleon that moves in the nuclear mean field created by the even-even core.
We further assume that the nuclear ground state may be statically deformed. 
Dynamical deformations 
are assumed to be small, and $E_{rot}\ll E_{vib}$, i.e., the adiabatic assumption holds for the separation of the rotational and vibrational motion for even-even and odd-$A$ targets. Under those assumptions, the nuclear Hamiltonian can be written as,
\begin{equation}
H=H_{rot}+H_{vib}+H_{p}+H_{int},  \label{hamiltonian}
\end{equation}%
where $H_{rot}$ is the rotational energy operator; $H_{vib}$ is the
vibrational energy operator; $H_{p}$ is the energy operator of the unpaired
nucleon (single-particle operator); and $H_{int}$ is the interaction energy
operator of the unpaired nucleon with the nuclear even-even core field. \
The corresponding Schr\"odinger equation can be written as $H\Psi =E\Psi $,
where the nuclear eigenfunction $\Psi $ describes single-particle, vibrational and
rotational motions.

\subsection{Even-even nuclei}

Following our assumptions for actinides, we do not consider unpaired
nucleons for even-even nuclei (nor single-particle excitations), therefore
the nuclear Hamiltonian $H\equiv H_{rot}+H_{vib}$ describes the collective
motion only, neglecting the interaction of vibrational and rotational
states. Adiabatic approximation for the collective motion means that the
even-even nuclear wave function can be factorized into a rotational and a
vibrational part $\Psi =\Phi _{rot}(\Theta )\left\vert n(\lambda_{ph})\right\rangle $,
where the rotational wave function $\Phi_{rot}(\Theta )$ depends on the Euler
angles $\Theta $, and the vibrational wave function $\left\vert n(\lambda_{ph})\right\rangle$
depends on the number $n_{ph}$ of excited vibrational phonons of multipolarity $\lambda_{ph} $
and, implicitly, on the corresponding vibrational deformation variables $\beta_2$, $\beta_3$, $\gamma$, and $\eta$
(e.g., $n(\lambda_{ph}=2)$ represents quadrupolar axial and non-axial phonons and
depends on $\beta$ and $\gamma$ quadrupolar deformations, respectively).
The phonon parity is $\pi _{ph}=(-1)^{\lambda _{ph}}$.
In this work we consider only one-phonon vibrational states.
The phonon description of vibrational states also implies that inter-band
transitions when the number of phonons changes by more than one ($\Delta
n_{ph}>1$) are forbidden (this is a well known selection rule of the
vibrational phonon model). Therefore, strongest inter-band transitions
occur between the bands built on one-phonon states and the ground state band.

Assuming that the nucleus behaves as a tri-axial rotor and denoting $K$ as the total angular momentum projection on the symmetry axis and $I$ $(I\geq K)$ as the total angular momentum, then the nuclear wave function $\Psi $ is mixed in $K$, and can be written in the form \cite{Bohr52,Preston75}
\begin{equation}
\Psi (IM\tau \Theta) = \sum_{K=0}^{I}~A_{K}^{I\tau }\left[ \frac{2I+1}{16\pi
^{2}(1+\delta _{K0})}\right] ^{1/2}\left[ D_{MK}^{I}(\theta ^{\prime
},\varphi ^{\prime })+(-1)^{I+\lambda _{ph}}D_{M-K}^{I}(\theta ^{\prime
},\varphi ^{\prime })\right] \left\vert n(\lambda _{ph})\right\rangle .
\label{wave-even}
\end{equation}%
The coefficients $A_{K}^{I\tau }$ are the $K$-mixing coefficients that depend
on all deformation parameters and satisfy orthonormality conditions
\begin{equation}
\sum_{K\geq 0}A_{K}^{I\tau }A_{K}^{I\tau ^{\prime }}=\delta _{\tau \tau
^{\prime }}\text{ \ \ and \ }\sum_{\tau }A_{K}^{I\tau }A_{K^{\prime }}^{I\tau
}=\delta _{KK^{\prime }},
\end{equation}%
\noindent where $\tau $ denotes the additional quantum numbers.
The expression \eqref{wave-even} can be considered a generalization of the wave
function given in Eq.~(91) of Ref.~\cite{Bohr52} and Eqs.~(9-59) and (9-60)
of Ref.~\cite{Preston75} (all restricted to quadrupolar deformations) for
any phonon multipolarity. The phonon wave function $n(\lambda _{ph})$ has been ignored
in those references, since for even-even nuclei the lowest-lying states are given by rotational
and vibrational modes with intrinsic spin 0 \cite{Preston75}. Note that if
$K=0$, then the condition $I+\lambda _{ph}=even$ holds and the phase factor becomes unity.
The phonon parity operator $\pi _{ph}=(-1)^{\lambda _{ph}}$ in Eq.~\eqref{wave-even} is
missing in Eq.~(3) of Ref.~\cite{Hecht62} as only quadrupolar vibrations
featuring positive parity were considered. In this work, the phonon parity
has been considered as needed for the proper inclusion of
both quadrupolar $\lambda _{ph}=2$ and octupolar $\lambda _{ph}=3$ vibrations.

The coefficients $A_{K}^{I\tau }$, which determine $K$-mixing in a
given rotational state due to the non-axiality, are to be determined from
the solution of the rotational problem, and can be obtained as described in
Ref.~\cite{SHEMMAN} by solving the Schr\"odinger equation corresponding to
the non-axial soft nuclear Hamiltonian. The code SHEMMAN adjusts the
soft-rotator Hamiltonian parameters by comparing the energies of the
experimental and calculated collective levels. Therefore, the coefficients $%
A_{K}^{I\tau }$ are determined independently from coupled-channels
equations.

For axially-symmetric nuclei the projection $K$ becomes a good
quantum number and only one $A_{K}^{I\tau }$ coefficient is different from zero.
Therefore, the sum over $K$ and the $K$-mixing disappears in Eq.~\eqref{wave-even}, and we obtain,
\begin{equation}
\Psi (IMK \lambda _{ph}\Theta ) = \left[ \frac{2I+1}{16\pi ^{2}(1+\delta _{K0})}\right] ^{1/2}\left[
D_{MK}^{I}(\theta ^{\prime },\varphi ^{\prime })+(-1)^{I+\lambda_{ph}}
D_{M-K}^{I}(\theta ^{\prime },\varphi ^{\prime }) \right] \left\vert n(\lambda _{ph})\right\rangle .
\label{wave-even-sym}
\end{equation}

The optical model calculations and fitting undertaken in this work assumed that even-even nuclei
were axially symmetric rotors in their ground state. However, non-axial formulation of the wave
function given by Eq.~\eqref{wave-even} was retained for the calculation of most general coupling matrix elements.

\subsection{Odd-$A$ nuclei}

Following our assumptions for odd actinides, we neglect vibrational excitations of the core. The total angular momentum of the nucleus $\overrightarrow{I}$, a constant of the motion, is now the sum of two parts, $\overrightarrow{C}$, the even-even core's angular momentum, and $\overrightarrow{j}$, the nucleon angular momentum, where $\overrightarrow{I}=\overrightarrow{C}+\overrightarrow{j}$. It is customary to take the projections of $\overrightarrow{I}$ and $\overrightarrow{j}$along the intrinsic $z^\prime$-axis to be $K$ and $\Omega $, respectively, while the $z$-component of $\overrightarrow{I}$ is $M$. Since the potential in which the unpaired nucleon moves is, in general, neither spherically nor axially symmetric, $j^{2}$, $K$, and $\Omega $, are not constants of the motion. Following Refs.~\cite{Hecht62,Preston75}, we assume the extreme extracore single-particle model for the Hamiltonian $H_{p}$ describing the unpaired nucleon. The nucleon single-particle wavefunction $\chi _{\nu }$ is the eigenfunction of the single-particle Hamiltonian $H_{p}$\ that describes a nucleon moving in the deformed nuclear mean field (e.g., Nilsson or Woods-Saxon potential), i.e., $H_{p}\chi _{\nu }=E_{\nu}\chi _{\nu }$, being $E_{\nu}$ the single-particle energy corresponding to the single-particle state $\nu$.

Following Eq.~(3) of reference \cite{Hecht62} the nuclear wave function in the extreme single-particle model is given by
\begin{align}
\Psi (IM\tau \Theta )&=\left[ \frac{2I+1}{16\pi ^{2}}\right] ^{1/2} {\sum_{K>0}}^{\prime} C_{K }^{I\tau} \left[ D_{MK}^{I}(\theta^{\prime },\varphi ^{\prime })\chi _{\nu }+(-1)^{I-1/2}D_{M-K}^{I}(\theta^{\prime },\varphi ^{\prime })\pi _{\chi }\chi _{-\nu }\right], \label{wave-odd-gen}
\end{align}%
where $\pi _{\chi }$ is the parity of the intrinsic wave function, and $C_{K }^{I\tau}$ are the $K$-mixing coefficients that depend on deformation parameters, and are determined from the solution of the rotational problem (e.g., see Refs.~\cite{davidov60,davidov62}). In the asymmetric case ellipsoidal symmetry imposes additional restrictions on quantum numbers $K$ and $\Omega$ (that characterizes the single-particle wave function $\chi_\nu$ -- see Appendix~2), namely that ($K-\Omega$) must be an even integer \cite{Hecht62}. This restriction on $K$ values is indicated by the apostrophe on the summation symbol ${\sum_{K>0}}^{\prime}$ in Eq.~\eqref{wave-odd-gen}.

It is remarkable that the odd-$A$ nucleus wave function given by Eq.~\eqref{wave-odd-gen} has exactly the same collective angular operator structure (i.e., Wigner functions) that the even-nucleus wave function given by Eq.~\eqref{wave-even}. Such analogy allows using the same matrix element derived for the even-even nuclear wave function to obtain the odd-$A$ matrix element as will be shown in Appendix~2. There are only two differences regarding the odd and even-even nuclear wave functions: 1) the phase in front of the second Wigner function changes to $(-1)^{I-1/2}\pi_{\chi }$ (odd-$A$ wave function) from the even-even case phase $(-1)^{I+\lambda _{ph}}$; and 2) there is a single particle wave function $\chi_\nu$ ($\chi_{-\nu}$) that multiplies the first (second) Wigner functions in the odd-$A$ case in Eq.~\eqref{wave-odd-gen}.

For axially-symmetric nuclei the sum over $K$ disappears, $K$ becomes a good quantum number that labels the single-particle state $\nu$ (i.e., $\chi_\nu\equiv\chi_K$), which automatically leads to Eq.~(9-139) of Ref.~\cite{Preston75}:
\begin{equation}
\Psi (IMK \Theta ) = \left[ \frac{2I+1}{16\pi ^{2}}\right] ^{1/2}\left[D_{MK}^{I}(\theta ^{\prime },\varphi ^{\prime })\chi _{K}+(-1)^{I-1/2}
D_{M-K}^{I}(\theta ^{\prime },\varphi ^{\prime }) \pi _{\chi} \chi _{-K}\right].  \label{wave-odd-sym}
\end{equation}

Note that axial symmetry is an acceptable approximation for odd-$A$ nuclei, as demonstrated by an extensive use of the rigid-rotor assumption in odd-actinide optical model potentials.

\section{Coupled-channels matrix elements for extended coupling scheme}

Starting from the derived potential multipolar expansion of the optical model potential (see eq.~\eqref{V_multip_exp} in Appendix~1), and using nuclear wave functions given above, we can derive the general expressions \eqref{matrix-element} and \eqref{matrix-element-odd} for even-even and odd targets, correspondingly (see Appendix~2 for a detailed derivation) for coupling matrix elements $\left\langle i\right\vert V(r,\theta ,\varphi )\left\vert f\right\rangle$ between nuclear states $\left\vert i\right\rangle$ and $\left\vert f\right\rangle$. In those expressions, the reduced matrix elements have to be defined, and the radial functions are given by Eq.~\eqref{eq:vlambda}, which for $i=1$ are being equal to the radial functions used in matrix elements of the conventional rigid-rotor potential.

Reduced matrix elements entering Eqs.~\eqref{matrix-element} and ~\eqref{matrix-element-odd} are different for couplings in even-even and odd-$A$ nuclei as the corresponding nuclear wave functions given by Eqs.~\eqref{wave-even}--\eqref{wave-odd-sym} are different. Ready to use expressions for calculations of the reduced matrix elements are given by Eq.~\eqref{red-mat-even-gen} for even-even targets, and by Eq.~\eqref{red-mat-odd-gen} for odd-$A$ targets. The reduced matrix elements also define the selection rules determining couplings, which are discussed in Appendix~2.

\begin{figure*}[t]
\vspace{-0.50cm}
\begin{center}
\includegraphics[width=1.17\textwidth]{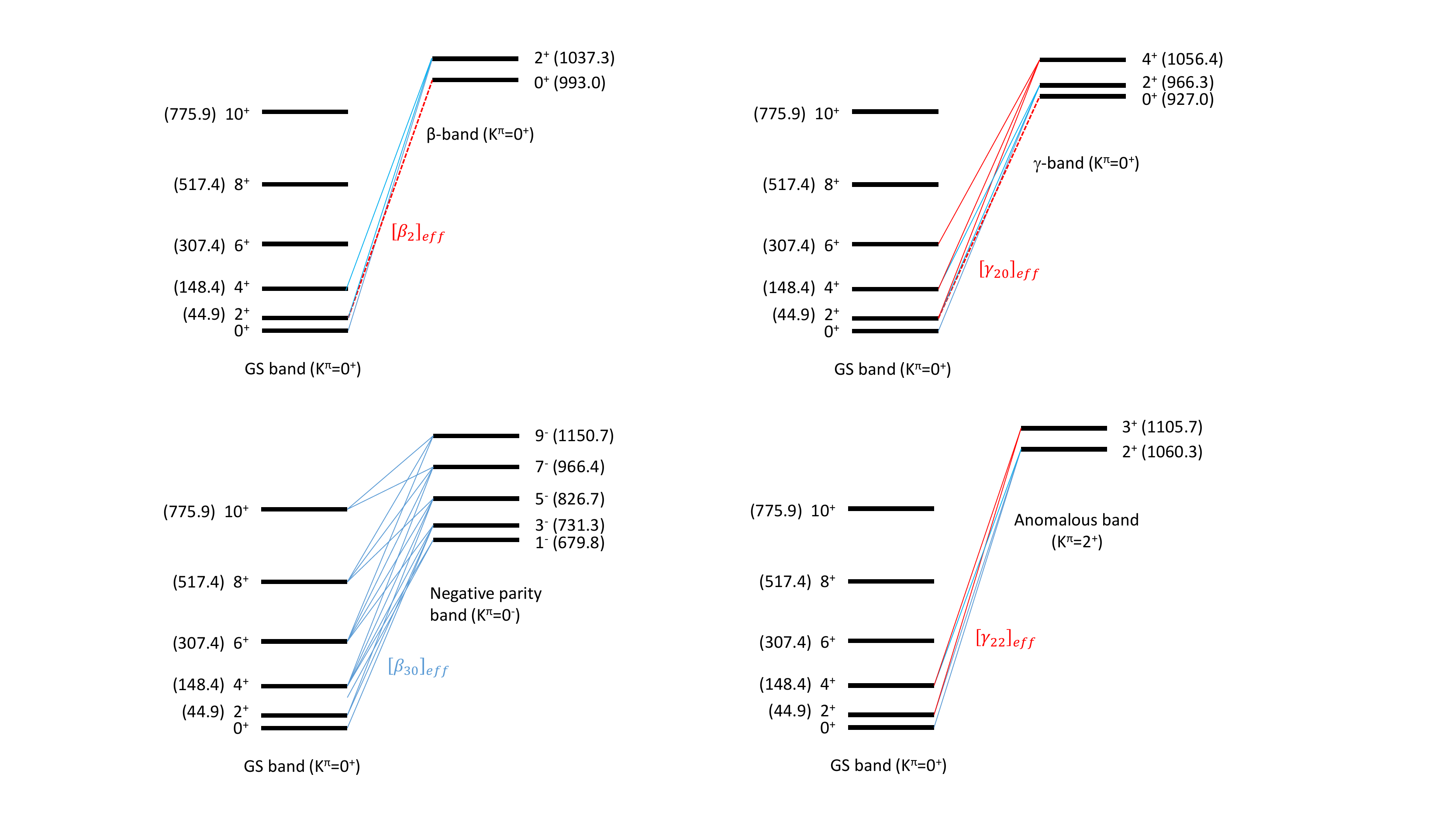}
\end{center}
\vspace{-1.cm}
\caption{(Color online) Multi-band coupling scheme (not to scale) is shown for a generalized optical model description of $n$+$^{238}$U reaction. Transitions between excited bands and the ground state band are indicated, those between excited vibrational bands are forbidden within the vibrational phonon model. Numbers in parentheses next to the level spin are corresponding level excitation energies in keV. The trivial in-band quadrupolar transitions with $\Delta K=0$ are omitted. The selection rules are discussed in Appendix~2. 18 levels are coupled for neutron induced reactions including the ground state rotational band ($K=0^+$), and four excited vibrational bands ($K=0^+$ $\beta$-band, $K=0^+$ $\gamma-$band,  $K=0^-$ octupole band, and anomalous $K=2^+$ non-axial band). For proton induced reactions the coupling from the ground state band to isobar-analog states (IAS) should be added.}
\label{fig:coupling}
\end{figure*}

The general expressions mentioned above constitute an important result of this work allowing the extension of Tamura's coupled-channels formalism to describe the coupling of additional excited bands both for even-even and odd-$A$ targets.

An example of the coupling scheme employed in the current work is shown in Fig.~\ref{fig:coupling} for a neutron induced reaction on $^{238}$U target. The corresponding effective deformations (e.g., $[\beta_2]_{eff}$) that define the inter-band coupling strength are shown near the transitions. The highest-energy state considered in the ground state rotational band of $^{238}$U
was the $10^+$ located at 775.9~keV; the rigid rotor assumption breaks down above that excitation energy (level energies do not follow the rigid rotor prediction $E(I)\sim I(I+1)$).

\section{Dispersive optical model formalism}

Pioneering dispersive optical model analyses for nucleon scattering have
been carried out by Lipperheide \cite{li66,li67}, Passatore \cite{pa67} and
Lipperheide and Schmidt \cite{lisc68}. Both bound and scattering states were
calculated using the same nuclear mean field constrained by dispersive
relations \cite%
{mang84,mangsa86,masa86,masa87np,johoma87,masa88,masa89,fiwida89,dewara89,tochde90, masa91,masa91rev,moro04}.
Recent works to improve the description of the bound states using the
dispersive optical model, and describe reactions on unstable targets off the
$\beta $-stability valley have been published by a group at Washington
University, St. Louis, Missouri (see Ref.~\cite{di10} and references therein).
Additional constraint imposed by dispersion
relations helps to reduce the ambiguities in deriving phenomenological OMP
parameters from the experimental data, and effectively also reduces the
number of phenomenological parameters.

However, a few studies have been devoted to derive dispersive optical model
parameterizations for strongly deformed nuclei, where coupled-channels
formalism should include 
potential corrections arising from
dispersion effects \cite{Merchant:92,rode97,sm01,sm02,sm04}. Recently we
derived an isospin dependent optical model potentials for actinides based on
rigid-rotor couplings \cite{casoqu05,Capote:08}. We will summarize
below the formalism employed in this work.

Assuming that the geometry of the imaginary terms of the OMP is
energy-independent, then the deformed optical model potential for incident
nucleons may be written as
\begin{align}
V(r,R(\theta ,\varphi ),E)& =-V_{HF}(E)f_{_{WS}}(r,R_{HF}(\theta ,\varphi ))
\notag \\
& -\left[ \Delta V_{v}(E)+iW_{v}(E)\right] f_{_{WS}}(r,R_{v}(\theta ,\varphi
))  \notag \\
& -\left[ \Delta V_{s}(E)+iW_{s}(E)\right] g_{_{WS}}(r,R_{s}(\theta ,\varphi
))  \notag \\
& +\left( \frac{\hbar }{m_{\pi }c}\right) ^{2}\left[ V_{so}(E)+\Delta
V_{so}(E)+iW_{so}(E)\right] \times \frac{1}{r}\frac{d}{dr}%
f_{_{WS}}(r,R_{so})(\hat{l}\cdot \hat{\sigma})  \label{OMPfull} \\
& +V_{Coul}(r,R_{c}(\theta ,\varphi ))  \notag
\end{align}%
where the first term is the real smooth volume potential $V_{HF}(E)$.
Successive complex-valued terms are the volume, surface, and spin-orbit
potentials, all containing the corresponding 
dispersive contributions $\Delta V_{v}(E),\Delta V_{s}(E)$ and $\Delta V_{so}(E)$
discussed in section \ref{Dispersive relations}. The Coulomb potential
term $V_{Coul}$ is needed for incident protons. We followed our previous
studies \cite{socaqu05,casoqu05,Capote:08} and connected the imaginary spin-orbit
potential $W_{so}(E)$ to the real spin orbit potential $V_{so}(E)$ by a
dispersion relation as discussed by Walter \cite{wa96}.

The geometrical Woods-Saxon form factors are given as \
\begin{equation}
f_{_{WS}}(r,R_{i}(\theta ,\varphi ))~=~\left[ 1+\exp [\left( r-R_{i}(\theta
,\varphi )\right) /a_{i}]\right] ^{-1},\qquad i=HF,v
\end{equation}%
\begin{equation*}
g_{_{WS}}(r,R_{s}(\theta ,\varphi ))=-4a_{s}\frac{d}{dr}f(r,R_{s}(\theta
,\varphi ))
\end{equation*}%
\begin{equation*}
f_{_{WS}}(r,R_{0,so})~=~\left[ 1+\exp [(r-R_{0,so})/a_{i}]\right] ^{-1},
\end{equation*}%
where deformed radii $R_{i}(\theta ,\varphi )$ are described by Eq.~%
\eqref{radius}. An spherical spin-orbit potential was used with constant
radius equal to $R_{0,so}$. The deformed Coulomb potential $%
V_{Coul}(r,R_{c}(\theta ,\varphi ))$ was calculated using a multipole
expansion of charged ellipsoid with a uniform charge density within the
Coulomb radius $R_{0,C}$ and zero outside as suggested by Bassel \etal\cite{badrsa62}.
The spherical term of the Coulomb potential was
calculated by taking account of the diffuseness of the charge density
distribution of the form $f_{c}=\left[ 1+\exp \left( r-R_{0,C}\right)
/a_{c}\right] ^{-1}$ \cite{chiwya97}.

The incident nucleon energy $E=E_{n}$ is equal to the incident energy for
neutrons; for incident protons we have to take into account the modification
of the nuclear potential by Coulomb repulsion . We assume that the
\ql effective\qr interacting energy of the proton is $E=E_{p}-C_{Coul}\frac{%
Z^{\prime }}{A^{1/3}}$, being $A$, $Z^{\prime }$ the target mass and atomic
numbers.
The term $C_{Coul}\frac{Z^{\prime }}{A^{1/3}}$ is an estimate of the kinetic
energy loss of the incident proton in the interaction region due to Coulomb
repulsion. The use of the \ql effective\qr interacting energy of the proton is a
generalization of the previously used Coulomb corrections, which considers
such corrections in all orders. The constant $C_{Coul}$ is an adjustable
constant meant to account for the "effective" radius of interaction of
proton in nucleus. Avoiding the use of Coulomb
corrections is a pre-condition to the Lane consistency, otherwise the
resulting potential is not symmetric with respect to the nucleon charge.

The present optical potential includes relativistic corrections as discussed
by Elton \cite{el66}. Firstly, the nucleon wave number $k$ was calculated in
the relativistic form $(\hbar k)^{2}=[E^{2}-(M_{p}c^{2})^{2}]/c^{2}$ where $%
E $ denotes the total energy of projectile, $M_{p}$ the projectile rest
mass, and $c$ the light velocity. Secondly, projectile and target masses
were replaced by corresponding relativistic energies in reduced mass
formulae. However, the change of the potential depth related to the
transformation from the Dirac equation was not considered in this work.

\subsection{Dispersive relations\label{Dispersive relations}}

In a dispersion relation treatment, the real potential strength consists of
a term which varies slowly with energy, the so called Hartree-Fock (HF)
term, $V_{HF}(E)$, plus a dynamic (polarization) term, $\triangle V(E)$,
which is calculated using a dispersion relation. Under favorable conditions
of analyticity in the complex $E$-plane the real part $\Delta V(E)$ can be
constructed from the knowledge of the imaginary part $W(E)$ on the real axis
through the dispersion relation%
\begin{equation}
\Delta V(E)=\frac{\mathcal{P}}{\pi }{\int_{-\infty }^{\infty }}\frac{%
W(E^{\prime })}{E^{\prime }-E}dE^{\prime }~,  \label{disp_integral}
\end{equation}%
where $\mathcal{P}${\ means that the principal value of the integral should
be taken.} Assuming that $\Delta V(E=E_{_{F}})=0$, where $E_{_{F}}$ is the
Fermi energy, Eq.~\eqref{disp_integral} can also be be written in the
subtracted form
\begin{equation}
\Delta V(E)={\frac{\mathcal{P}}{\pi }\int_{-\infty }^{\infty }{W(E^{\prime })%
}\left( \frac{1}{E^{\prime }-E}-\frac{1}{E^{\prime }-E_{F}}\right)
dE^{\prime }}~.  \label{integral_subs}
\end{equation}%
Here $E_{F}$ denotes the Fermi energy, determined as $E_{F}(Z,A)=-\frac{1}{2}%
\left[ S_{n}(Z,A)+S_{n}(Z,A+1)\right] $ for neutrons and $E_{F}(Z,A)=-\frac{1%
}{2}\left[ S_{p}(Z,A)+S_{p}(Z+1,A+1)\right] $ for protons, where $S_{i}(Z,A)$
denotes the separation energy of nucleon $i$ from a nucleus labeled by $Z$
and $A.$ The symmetry condition $W(2E_{F}-E)=W(E)$ (for symmetric imaginary
potentials) is used to extend the imaginary part of the OMP for energies
below the Fermi energy, allowing for the calculation of the dispersive
integral.

It is known that the energy dependence of the depth $V_{HF}(E)$ is due to the replacement of a microscopic nonlocal HF potential by a local equivalent \cite{pebu62}. For a Gaussian non-locality $V_{HF}(E)$ is a linear function of $E$ for large negative $E$ and is an exponential for large positive $E$. Following Mahaux and Sartor \cite{masa91}, the energy dependence of the smooth \ql Hartree-Fock\qr part of the nuclear mean field is taken as that found by Lipperheide \cite{li67}:
\begin{equation}
V_{HF}(E)=A_{HF}\exp (-\lambda _{_{HF}}(E-E_{F}))  \label{VHF}
\end{equation}%
where the parameters $A_{HF}$ and $\lambda _{_{HF}}$ are undetermined
constants. Eq.~\eqref{VHF} can be used to describe the HF potential in
the scattering regime \cite{masa91} for $E>0$.

It is useful to represent the variation of surface $W_{s}(E)$ and volume absorption potential $W_{v}(E)$ depth with energy in functional forms suitable for the dispersive optical model analysis, which are integrable analytically \cite{PRC-disp}. An energy dependence for the imaginary volume term has been suggested in studies of nuclear matter theory by Brown and Rho \cite{brrh81}:
\begin{equation}
W_{v}(E)=A_{v}~\frac{(E-E_{F})^{2}}{(E-E_{F})^{2}+(B_{v})^{2}}  \label{WV3}
\end{equation}
where $A_{v}$ and $B_{v}$ are undetermined constants. An energy dependence
for the imaginary-surface term has been suggested by Delaroche \etal\cite{dewara89} to be:
\begin{equation}
W_{s}(E)=A_{s}~\frac{(E-E_{F})^{2}}{(E-E_{F})^{2}+(B_{s})^{2}}~\exp
(-C_{s}\vert E-E_{F}\vert )  \label{WS3}
\end{equation}
where $A_{s},B_{s}$ and $C_{s}$ are undetermined constants.

The isospin dependence of the potential (the Lane term \cite{la62a,la62b})
was considered in real $V_{HF}(E)$ and imaginary surface $W_{s}(E)$
potentials as follow,
\begin{align}
A_{HF}& =V_{0}\left[ 1+(-1)^{Z^{\prime }+1}\frac{C_{viso}}{V_{0}}\frac{N-Z}{A%
}\right]  \label{AHF} \\
A_{s}& =W_{0}\left[ 1+(-1)^{Z^{\prime }+1}\frac{C_{wiso}}{W_{0}}\frac{N-Z}{A}%
\right]
\end{align}%
where $V_{0},C_{viso},W_{0}$ and $C_{wiso}$ are undetermined constants. Many
authors found that the imaginary volume potential does not depend on the
isospin.

For the energy dependence of the spherical spin-orbit potential we used the functional form suggested by Koning and Delaroche \cite{kode03}, which is convenient for the calculation of the dispersive contribution \cite{PRC-disp}, namely:
\begin{align}
V_{so}(E)& =V_{SO}\exp (-\lambda _{so}(E-E_{F}))\text{ }
\label{spin-orbit-real} \\
W_{so}(E)& =W_{SO}~\frac{(E-E_{F})^{2}}{(E-E_{F})^{2}+(B_{so})^{2}}
\label{spin-orbit-imag}
\end{align}
where $V_{SO},\lambda _{so},W_{SO}$ and $B_{so}$ are undetermined constants.

\subsection{High energy behavior of the volume absorption}

The DOM analysis of neutron scattering on $^{27}$Al \cite{mocaqu02} showed the importance of the dispersive contribution to describe ${\sigma }_{T}(E)$ data for energies above 100 MeV using a non-symmetric version of the volume absorptive potential for large positive and large negative energies as proposed by Mahaux and Sartor \cite{masa91}. Similar behavior was confirmed in $^{232}$Th dispersive coupled-channels analysis of nucleon induced reactions \cite{socaqu05} and in combined dispersive coupled-channels analysis of nucleon induced reactions on $^{232}$Th and $^{238}$U \cite{casoqu05,Capote:08}. We use the same formalism in this work, which is briefly described below.

Following Mahaux and Sartor \cite{masa91}, the assumption that the imaginary potential $W_{v}(E)$ is symmetric about $E^{\prime }=E_{F}$ (according to equation $W(2E_{F}-E)=W(E)$) is modified above some fixed energy $E_{a}$, which is expected to be close to 60 MeV. However this value is fairly arbitrary \cite{masa91} and we will use it as a fitting parameter. Let us assume the imaginary potential to be used in the dispersive integral is denoted by $\widetilde{W}_{v}(E)$, then we can write \cite{masa91rev}
\begin{equation}
\widetilde{W}_{v}(E)=W_{v}(E)-W_{v}(E)\frac{(E_{F}-E-E_{a})^{2}}{%
(E_{F}-E-E_{a})^{2}+E_{a}^{2}}\text{, for }E<E_{F}-E_{a}\text{ (bound regime)%
},  \label{WVnonlocal1}
\end{equation}%
and
\begin{equation}
\widetilde{W}_{v}(E)=W_{v}(E)+\alpha_v \left[ \sqrt{E}+\frac{%
(E_{F}+E_{a})^{3/2}}{2E}-\frac{3}{2}\sqrt{(E_{F}+E_{a})}\right]\text{, for }%
E>E_{F}+E_{a}\text{ (scattering regime)}.  \label{WVnonlocal2}
\end{equation}%
These functional forms are chosen in such a way that the function and its first derivative are continuous at $E^{\prime }=\left\vert E_{F}-E_{a}\right\vert $. At large positive energies nucleons sense the \textquotedblright hard core\textquotedblright\ repulsive region of the nucleon-nucleon interaction and $\widetilde{W}_{v}(E)$ diverges like $\alpha _v
\sqrt{E}$. Using a model of a dilute Fermi gas hard-sphere the coefficient $\alpha _v$ can be estimated to be equal to 1.65 MeV$^{1/2}$ \cite{masa86}, but the actual value is a model parameter.
On the contrary, at large negative energies the volume absorption decreases and goes asymptotically to zero. The asymmetric form of the volume imaginary potential of Eqs.~\eqref{WVnonlocal1} and \eqref{WVnonlocal2} results in a dispersion relation that must be calculated directly from Eqs.~\eqref{disp_integral} and \eqref{integral_subs}, and separates into three additive terms \cite{vaweto00,CPC}.
Therefore, we write the dispersive volume correction in the form
\begin{equation}
\triangle \widetilde{V}_{v}(E)=\triangle V_{v}(E)+\triangle
V_{<}(E)+ \alpha_v \triangle V_{>}(E),
\end{equation}%
where $\triangle V_{v}(E)$ is the dispersive correction due to the symmetric imaginary potential of Eq.~\eqref{WV3} which is calculated following Ref.~\cite{PRC-disp}, and the terms $\triangle V_{<}(E) $ and $\triangle V_{>}(E)$ are the dispersive corrections due to the asymmetric terms of Eqs.~\eqref{WVnonlocal1} and \eqref{WVnonlocal2}, respectively calculated as described in Ref.~\cite{CPC}. Unfortunately, the equation giving the dispersive correction $\triangle V_{>}(E)$  in Ref.~\cite{CPC} contains several typos, so we give the correct expression below.

Using the notation $E_{_L}=E_{_F}+E_a$, the $\Delta V_{>}(E)$ contribution for volume nonlocality correction is exactly:
\begin{eqnarray}
\Delta V_{>}(E)=&&\frac 1\pi \times \left[ \sqrt{\vert E_{_F}\vert }\arctan \frac{2%
\sqrt{E_{_L}\vert E_{_F}\vert }}{E_{_L}-\vert E_{_F}\vert }+\frac{E_{_L}^{3/2}}{2E_{_F}}ln\frac{%
E_a}{E_{_L}}\right] +  \label{T2} \\
&&\frac 1\pi \times \left\{
\begin{array}{ll}
\sqrt{E}\ln \frac{\sqrt{E}+\sqrt{E_{_L}}}{\sqrt{E}-\sqrt{E_{_L}}}+\frac{3%
}2\sqrt{E_{_L}}\ln \frac{E-E_{_L}}{E_a}+\frac{E_{_L}^{3/2}}{2E}\ln \frac{%
E_{_L}}{E-E_{_L}} & \textrm{for }E>E_{_L}\\
\\
\frac{3}2\sqrt{E_{_L}}\ln \frac{2^{4/3}E_{_L}}{E_a} & \textrm{for }E=E_{_L} \B \\
\\
\sqrt{E}\ln \frac{\sqrt{E}+\sqrt{E_{_L}}}{\sqrt{E_{_L}}-\sqrt{E}}+\frac{3}2%
\sqrt{E_{_L}}\ln \frac{E_{_L}-E}{E_a}+\frac{E_{_L}^{3/2}}{2E}\ln \frac{%
E_{_L}}{E_{_L}-E} & \textrm{for }E_{_L}>E>0 \B \\
\\
\frac{3}2\sqrt{E_{_L}}\ln \frac{E_{_L}}{E_a}+\frac 12\sqrt{E_{_L}} & \textrm{%
for }E=0 \B \\
\\
-\sqrt{\vert E\vert  }\arctan \frac{2\sqrt{E_{_L}\left\vert  E\right\vert  }}{%
E_{_L}-\left\vert E\right\vert  }+\frac{3}2\sqrt{E_{_L}}\ln \frac{E_{_L}-E}{E_a}+%
\frac{E_{_L}^{3/2}}{2E}\ln \frac{E_{_L}}{E_{_L}-E} & \textrm{for }E<0\nonumber
\end{array}
\right.
\end{eqnarray}

The resulting dispersive correction for the asymmetric case starts to increase already for energies above 50 MeV, making
a significant contribution to the real part of the OMP at high energies. It should be noted that non-locality corrections (Eqs.~\eqref{WVnonlocal1} and \eqref{WVnonlocal2}) can be used either for the volume or surface imaginary potential; however, Mahaux and Sartor \cite{masa91} have shown that nonlocality consideration for the surface imaginary potential has a very small effect on the calculated cross sections. Therefore in this work we followed Ref.~\cite{masa91rev} and only considered the effects of nonlocality in the volume absorption.

\subsection{Dispersive coupled-channels optical model analysis}

A survey of the experimental data for nucleon interaction on actinide nuclei spanning from 0.001 to 150~MeV used in the current work coincide with the data used by Soukhovitskii and coworkers about 10~years ago \cite{Soukhovitskii:04}. The total cross section data considered cover all the critical energy points which are necessary to reveal the structure due to the Ramsauer effect and extends up to 150 MeV. The experimental database available for actinide nuclei other than $^{232}$Th and $^{238}$U is very scarce, especially in the high energy range; but some angular distributions of scattered neutron, and total cross section measurements are available for fissile actinides and were considered in the fitting. Quasi-elastic scattering ($p$,$n$) data on
$^{232}$Th and $^{238}$U were also considered in the fit to fix the isovector terms of the potential as described in Ref.~\cite{Quesada:2007}.

The starting values of static deformation parameters $\beta_2$, $\beta_4$, and $\beta_6$ were taken from FRDM deformations theoretically derived by M\"oller and Nix \cite{Moller:95}. Evaluated neutron strength functions for actinide nuclei, $S_{0}$ and $S_{1}$ and potential scattering radius $R^{\prime }$ \cite{ripl3,CSWEG1991} were used for fine-tuning deformation parameters together with measured ($p$,$p$') angular distributions on excited states close to the Coulomb barrier ($E_p>20$~MeV) for actinide nuclei of interest. The minimization procedure used the quantity $\chi ^{2}$ as described in previous
references \cite{socaqu05,casoqu05,Capote:08}.

Both direct and statistical processes contribute to nucleon-nucleus elastic scattering at these energies. However, according to our estimation, the statistical processes are not important above 3~MeV on actinides so we neglect them in the OMP derivation. The direct processes, increasingly dominant at higher energies, can be described by the optical model.

The customary coupled-channels calculations were performed by coupling the ground state $K^{\pi }$ rotational band with the selected members of vibrational (single-particle) bands as shown in Appendix~3 for studied actinides. Optical model code OPTMAN \cite{sochiw05,somoch04} was used for OMP parameter fitting. We were using symmetric surface and nonsymmetric volume imaginary absorptive potentials as described in previous section. The dispersion integrals were calculated using analytical solutions \cite{CPC,PRC-disp} of dispersion relations with Eq.~\eqref{T2} for the $\Delta V_{>}(E)$ contribution.

In our formulation of the OMP in Eq.~\eqref{OMPfull} the geometrical parameters of the \ql Hartree-Fock\qr potential $r_{HF}$ and $a_{HF}$ are in general different from geometrical parameters $r_{v},a_{v},r_{s},a_{s}$ of the volume and surface absorptive potentials; however the real and imaginary spin-orbit terms share the same geometric $r_{so}$ and $a_{so}$ parameters. Also the surface $\Delta V_{s}$ and volume $\Delta V_{v}$ dispersive corrections share the same geometry of corresponding imaginary potentials $W_{s}$ and $W_{v}$, respectively. The employed actinide optical potential energy dependence is very simple and we use energy-independent geometry and the same OMP parameters for both neutron and proton projectiles. A rather weak dependence of the dispersive potential geometry on mass number $A$ was previously observed \cite{casoqu05,Capote:08}. The fitted dispersive potential parameters are listed in Table \ref{table:disopt}. These parameters have to be complemented by a corresponding coupling scheme, ground state deformations, and coupling strength parameters (listed in Appendix~3, tables \ref{table:stat-deformations},
\ref{table:U8-deformations}, \ref{table:Th2-deformations}, \ref{table:Pu9-deformations}, \ref{table:U5-deformations}, and \ref{table:U3-deformations}) to become a complete OMP parameter set for studied actinides.

\begin{table}[!thb]
\vspace{-3mm}
\caption{Dispersive coupled-channels optical model potential parameters for nucleon induced reaction on actinides.}
\label{table:disopt}%
\begin{tabular}{c|c|c|c|c}
\hline\hline
                    & \textbf{VOLUME} & \textbf{SURFACE} & \textbf{SPIN-ORBIT} & \textbf{COULOMB} \\ \hline
\textbf{Real}       & $V_0=50.47+0.0292\:(A-238)$~MeV &  & $V_{SO}=6.1$~MeV & $C_{Coul}=1.36$~MeV \\
\textbf{potential}  & $\lambda _{HF}=0.00977$~MeV$^{-1}$ & -- & $\lambda _{so}=0.005$~MeV$^{-1}$ &  \\
\textbf{parameters} & $C_{viso}=17.3$~MeV &  &  &  \\ \hline
                    & $A_v=11.81$~MeV & $W_0=17.43$~MeV & $W_{SO}=-3.1$~MeV &  \\
\textbf{Imaginary}  & $B_v=81.81$~MeV & $B_s=10.57$~MeV & $B_{so}=160$~MeV &  \\
\textbf{potential}  & $E_a=55$~MeV & $C_s=0.01331$~MeV$^{-1}$ &  &  \\
\textbf{parameters} & $\alpha_v=0.355$~MeV$^{1/2}$ & $C_{wiso}=28.9$~MeV &  &  \\ \hline
                    & $r_{HF}=1.2468-0.00183\:(A-238)$~fm & $r_s=1.1717+0.0041\:(A-238)$~fm & $r_{so}=1.1214$~fm & $r_{c}=1.2894$~fm \\
\textbf{Potential}  & $a_{HF}=0.638+0.002134\:(A-238)$~fm & $a_s=0.618$~fm & $a_{so}=0.59$~fm    & $a_{c}=0.547$~fm \\
\textbf{geometry}   & $r_v=1.2657$~fm &  &  &  \\
                    & $a_v=0.6960-0.00021\:(A-238)$~fm &  &  &  \\ \hline\hline
\end{tabular}
\vspace{-4mm}
\end{table}

\section{Results and discussion}

\begin{table}[tbh]
\vspace{-0.25cm}
\caption{Comparison of calculated and experimental average resonance parameters. calculations using derived OMP are compared with results using the RIPL 2408 potential \cite{casoqu05,Capote:08}. The absolute uncertainties of evaluated strength functions and radii are given in parentheses.}
\label{table:average1}%
\begin{tabular}[t]{c|c|ccccc}
\hline\hline
Quantity & Reference & $^{232}$Th & $^{233}$U & $^{235}$U & $^{238}$U & $^{239}$Pu \T \\
\hline
& this work & 0.86 & 0.97 & 0.95 & 1.02 & 1.15 \\ 
$S_0$,(eV)$^{-1/2}10^{-4}$ & RIPL 2408 OMP \cite{casoqu05,Capote:08} & 0.86 & 0.93 & 0.93 & 0.92 & 1.18 \\ 
& RIPL-3 evaluation~\cite{ripl3} & 0.84 (.07) & 0.90 (.05) & 0.98 (.07) & 1.03 (.08) & 1.20 (.10) \\ 
& Porodzinski \etal\cite{posuma98} & 0.80 (.08) & 1.07 (.14) & 1.01 (.10) & 1.17 (.10) & 1.25 (.13) \\ 
& Mughabghab~2006~\cite{mu06} & 0.71 (.04) & 0.98 (.09) & 0.98 (.07) & 1.29 (.13) & 1.3 (.10) \\ \hline
& this work & 1.66 & 1.41 & 1.60 & 1.68 & 2.05 \\ 
$S_1$,(eV)$^{-1/2}10^{-4}$ &  RIPL 2408 OMP \cite{casoqu05,Capote:08} & 1.72 & 2.20 & 1.93 & 1.72 & 2.19 \\ 
& RIPL-3 evaluation~\cite{ripl3} & 1.50 (.30) & -- & -- & 1.60 (.20) & -- \\ 
& Mughabghab~2006~\cite{mu06} & 1.35 (.04) & -- & 1.8 (.30) & 2.17 (.19) & 2.30 (.40) \\ 
& CSEWG~1991~\cite{CSWEG1991} & 1.60 (.60) & -- & 1.8 (.30) & 1.70 (.30) & 2.30 (.40) \\ \hline
& this work & 9.68 & 9.60 & 9.58 & 9.52 & 9.50 \\ 
$R^{\prime }$,fm & RIPL 2408 OMP \cite{casoqu05,Capote:08} & 9.68 & 9.56 & 9.51 & 9.64 & 9.49 \\ 
& Mughabghab~2006~\cite{mu06} & 9.65 (.08) & 9.75 (.15) & 9.63 (.05) & 9.60 (.10) & 9.48 (.10) \\ 
& CSEWG~1991~\cite{CSWEG1991} & 9.65 (.30) & 9.75 (.20) & 9.65 (.10) & -- & 9.60 (.10) \\ \hline\hline
\end{tabular}%
\vspace{-0.20cm}
\end{table}

The derived dispersive potential parameters shown in Table \ref{table:disopt} for actinide nuclei combined with deformation parameters and coupling strengths tabulated in Appendix~3 were tested against selected experimental data. Average resonance parameters are consistently described by our calculations as can be seen in Table \ref{table:average1}. A similar level of agreement in calculated strength functions is seen for rigid-rotor OMP results (RIPL 2408) within quoted evaluated uncertainties. Current results for $S_0$ are  closer to evaluated ones for \UT~and \UF, but marginally worse for \PU. For $S_1$, calculations using the derived potential are closer to evaluations than those using the RIPL 2408 potential, but $S_1$ evaluated uncertainties are large.

The aim of this work is to improve the description of neutron inelastic scattering on actinides at incident neutron energies from keVs to a few MeVs. Unfortunately, many neutron-scattering physical observables (e.g., angular distributions of elastic and inelastic scattering, inelastic scattering cross sections, etc) measured on actinides for incident neutron energies below 3~MeV are influenced by a compound nuclear decay that can not be described by a coupled-channels optical model alone, which is a direct reaction model. Total cross section is one measured observable not affected by compound nuclear processes.

\begin{figure}[!tbph]
\vspace{-0.30cm}
\includegraphics[width=0.99\columnwidth,clip]{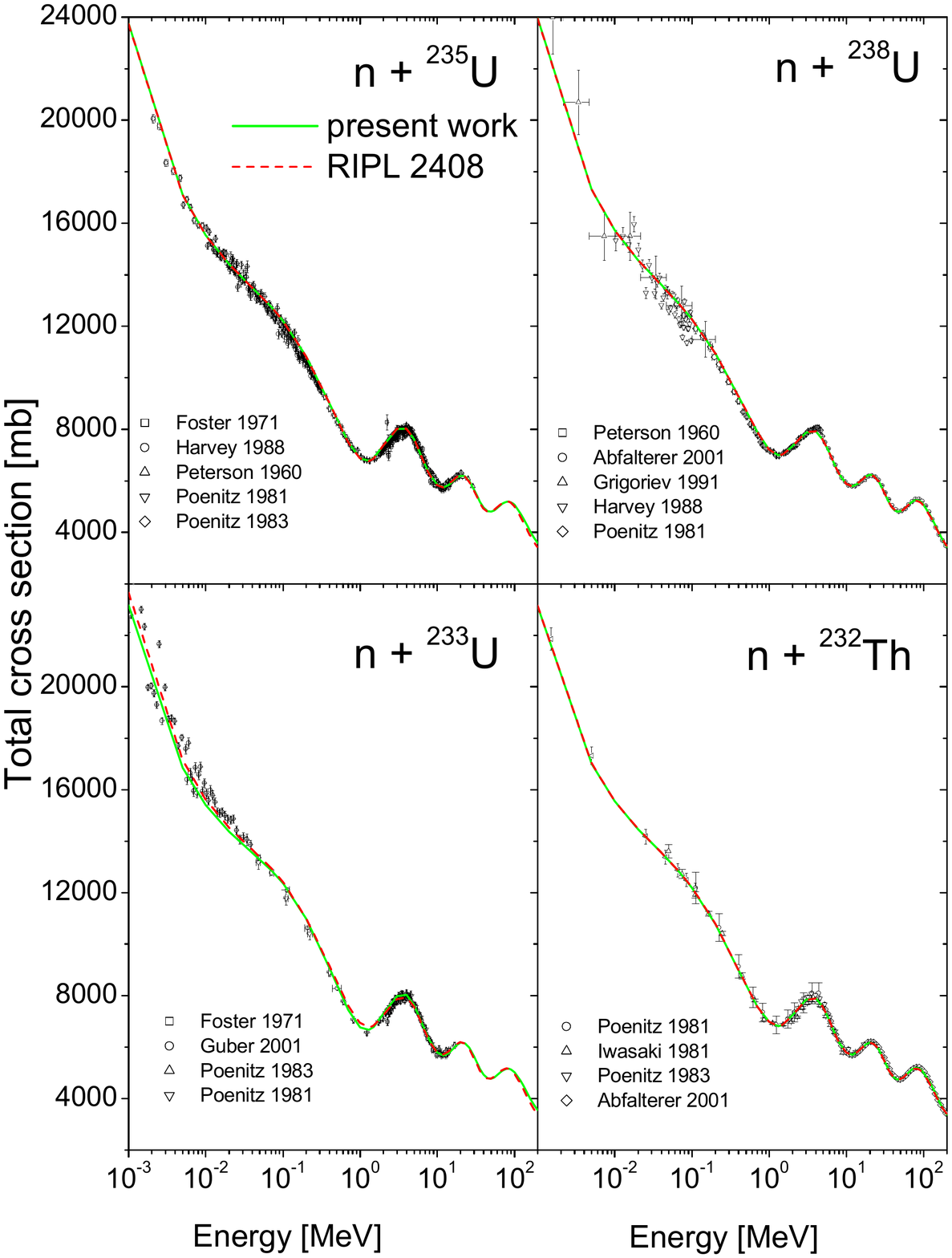}
\vspace{-1.3cm}
\caption{(Color online) Calculated total cross sections using the dispersive potential
with multiple-band coupling for neutron induced reactions on $^{233,235,238}$%
U and $^{232}$Th targets are compared with calculations using the RIPL 2408 potential \cite{casoqu05,Capote:08}. Experimental data are taken from Refs.~\cite{Foster:71,Harvey:88,Peterson:60,Poenitz:81,Poenitz:83,abbadi01,Guber:01,Iwasaki:81}.}
\label{fig:total}
\end{figure}

Total cross sections for $^{233,235,238}$U and $^{232}$Th nuclei were calculated with the derived potential parameters. An excellent agreement with data in the whole energy range is achieved as seen in Fig.~\ref{fig:total}. Similar agreement is also shown for the RIPL 2408 optical model potential based on a rigid-rotor structure. Calculated total cross sections do not show an improvement for a new potential as both calculated results are within quoted experimental uncertainty. However, it is remarkable that the newly derived potential with extended couplings preserves the achieved agreement with total cross section and other scattering observables shown in Refs.~\cite{casoqu05,Capote:08}, while allows predicting direct-interaction cross sections for additional coupled-levels in the whole energy range of interest.

From the physical point of view, a newly derived OMP utilizes a better description of the underlying nuclear structure compared to commonly used rigid-rotor potentials. The inadequacy of the rigid rotor description is clearly seen in the deviation from the $I(I+1)$ rule of the energies of excited states with large $I$ in the ground state rotational band of even-even actinides. E.g.,
for $^{238}$U and $^{232}$Th targets increasingly larger differences from the $I(I+1)$ rule can be seen
for levels with $I>8$ in the ground state rotational band, corresponding to excitation energies above $\simeq 500$ keV.
Such differences are mainly due to the change in moment of inertia (stretching of the soft nucleus). Additionally, $K$-mixing increases for higher-spin states 
as e.g., shown in a soft-rotator structure description \cite{SRM1,SRM2}. Therefore, the accuracy of the rigid-rotor description of the nuclear structure in even-even actinides rapidly deteriorates above 500~keV.

A lowest-energy vibrational mode in actinides is the octupolar vibration (e.g., the octupolar bandhead in $^{238}$U is located at 680~keV as shown in Fig.~\ref{fig:coupling}). This vibrational mode results in an enhanced neutron inelastic scattering cross section to negative parity levels $1^-$, $3^-$, $5^-$, ... of the low-lying octupolar band due to the strong band coupling. As a result the neutron inelastic scattering cross sections increases from 680~keV up to 1~MeV in the region of utmost importance for fast reactors. An smaller increase in inelastic scattering cross sections of actinides is also observed due to the strong coupling of the ground state band to other vibrational bands (e.g., $\beta$- and $\gamma$-vibrations). The OMP derived in this work offers a consistent description of the scattering of low energy neutrons on actinides that goes beyond the assumption of rigid rotor nuclear structure for those nuclei. However, the question is how to show a clear advantage of new potential using available experimental data?

It is generally accepted that differences of neutron total cross sections among neighboring nuclei provide an unusually
stringent test of optical models~\cite{shbera80,caphwh84,cadiph86,dianba03}. At the same time it has been shown that the standard optical model treatment fails to reproduce the observed differences of total cross sections for tungsten isotopes~\cite{dianba03}; incident neutron energies below 5 MeV are especially challenging due to the increasing impact of the target nuclear structure on calculated cross sections. Indeed, when the incident neutron energy becomes comparable with the energy of excited nuclear levels, then the weak-coupling assumption is not valid as the coupling becomes strong. In such conditions a commonly used DWBA method typically overestimates the observed scattering cross sections on vibrational levels as shown in Ref.~\cite{INDC0597}. Only a coupled-channels calculation with a consistent optical model potential would allow describing the observed cross section due to the strong coupling of vibrational bands.
\begin{figure*}[!bth]
\par
\begin{center}
\includegraphics[width=1.03\columnwidth,clip]{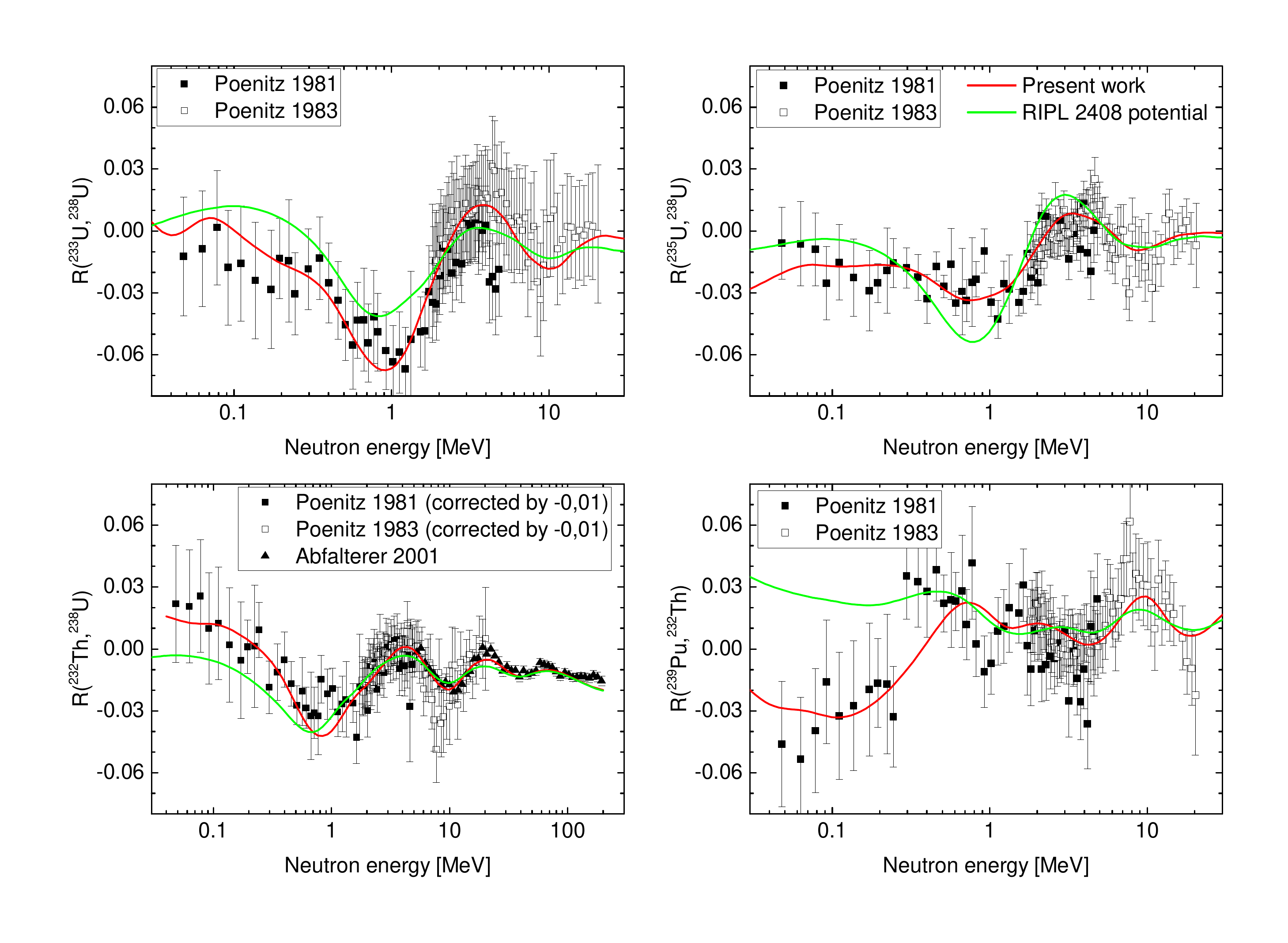}
\end{center}
\vspace{-1.cm}
\caption{(Color online) Energy dependence of the measured cross-section ratio
vs. calculated values using RIPL potential 2408 , and the potential derived in the current
work with parameters shown in Table \ref{table:disopt}
combined with deformation parameters tabulated in Appendix~3.}
\label{fig:ratio}
\vspace{-0.3cm}
\end{figure*}

We have already shown a good description of total cross sections in Fig.~\ref{fig:total}. However, a description of total cross-section differences is a much greater challenge. A natural test of the success of our approach is to check calculated total cross section differences for several pairs of actinides vs measured data at neutron incident energies below 5~MeV, where nuclear structure plays an important role. Energy-averaged total cross section data measured by Poenitz in 1981 and 1983~\cite{Poenitz:81,Poenitz:83} on $^{239}$Pu, $^{nat,238,235,233}$U and $^{232}$Th targets have been selected to benchmark our potential.
Poenitz high accuracy data goes from 50~keV neutron incident energy up to 20~MeV, and data were measured under the same conditions and at the same installation, thereby minimizing any uncertainty in the ratio. Energy-averaged total cross sections ${\sigma }_{tot}$ for $^{232}$Th and $^{238}$U nuclei measured by Abfalterer \etal\cite{abbadi01} from 5 to 200~MeV were also used to calculate the corresponding experimental ratio $R$ and associated experimental uncertainty. An small shift ($-0.01$) of Poenitz data was applied for the experimental $R(^{232}$Th,$^{238}$U$)$ ratio for a better matching of Abfalterer \etal\cite{abbadi01} data, well within the estimated uncertainty of the vertical scale (0.02~\cite{dianba03}) arising from uncertainties in the areal densities of the employed actinide targets.

Calculated total neutron cross section data for $^{232}$Th and $^{233,235,238}$U and $^{239}$Pu nuclei were used to obtain the energy-dependent ratio $R(A1,A2)$ of the total cross section difference $(\sigma_{tot}(A1)-\sigma_{tot}(A2))$ of targets $A1$ and $A2$ to the corresponding averaged total cross section $(\sigma_{tot}(A1)+\sigma_{tot}(A2))/2$ for the RIPL 2408 potential \cite{casoqu05,Capote:08} and the potential derived in this work. Calculated ratios $R$ are compared with experimental ratios derived from Poenitz and Abfalterer data in Fig.~\ref{fig:ratio} for the following target pairs: $R(^{239}$Pu,$^{232}$Th$)$, $R(^{233}$Pu,$^{238}$U$)$, $R(^{232}$Th,$^{238}$U$)$, and $R(^{235}$U,$^{238}$U$)$.
The measurements are very well reproduced for all target pairs by the dispersive OMP parameters shown in Table \ref{table:disopt} combined with the corresponding target deformation parameters and coupling strengths tabulated in Appendix~3. Therefore, the derived OMP for all five studied nuclei is validated. Rigid-rotor (RIPL 2408) potential \cite{casoqu05,Capote:08,ripl3} description of data is of inferior quality, especially for incident neutron energies below 3~MeV, where nuclear structure effects have the largest impact. Note that differences seen in the figure are typically around 2\% being the largest for $R(^{239}$Pu,$^{232}$Th$)$ at 50~keV (reaching about 5\%). Such differences are not noticeable in Fig.~\ref{fig:total}.

The depicted high quality description of data with derived potential at low incident neutron energies is due to the improved description of the nuclear structure of target nuclei considered in present calculations. The fact that such description was achieved with exactly the same regional potential for all targets with an energy-independent geometry, is an evidence of the predictive power of the dispersive optical model when combined with a proper description of the nuclear structure of target nuclei.

The elastic and inelastic angular distributions on selected ground-state band (GSB) coupled levels for the $n+^{238}$U reaction below 3.0~MeV were not included in the OMP fit as compound-nucleus contribution needs to be considered at those energies. A comprehensive calculation of $n+^{238}$U reactions was recently undertaken in Ref.~\cite{Capote:2014}; calculated angular distributions below 3~MeV were shown to be in very good agreement with existing experimental data. An equally good agreement was also shown for angular distributions of scattered neutrons (including contributions from the lowest levels of the ground state rotational band) measured at higher incident neutron energies from 4 up to 15~MeV~\cite{Capote:2014}.

\begin{figure*}[!tbh]
\par
\vspace{-0.50cm}
\begin{center}
\includegraphics[trim=6mm 6mm 0mm 0mm,width=1.03\columnwidth,clip]{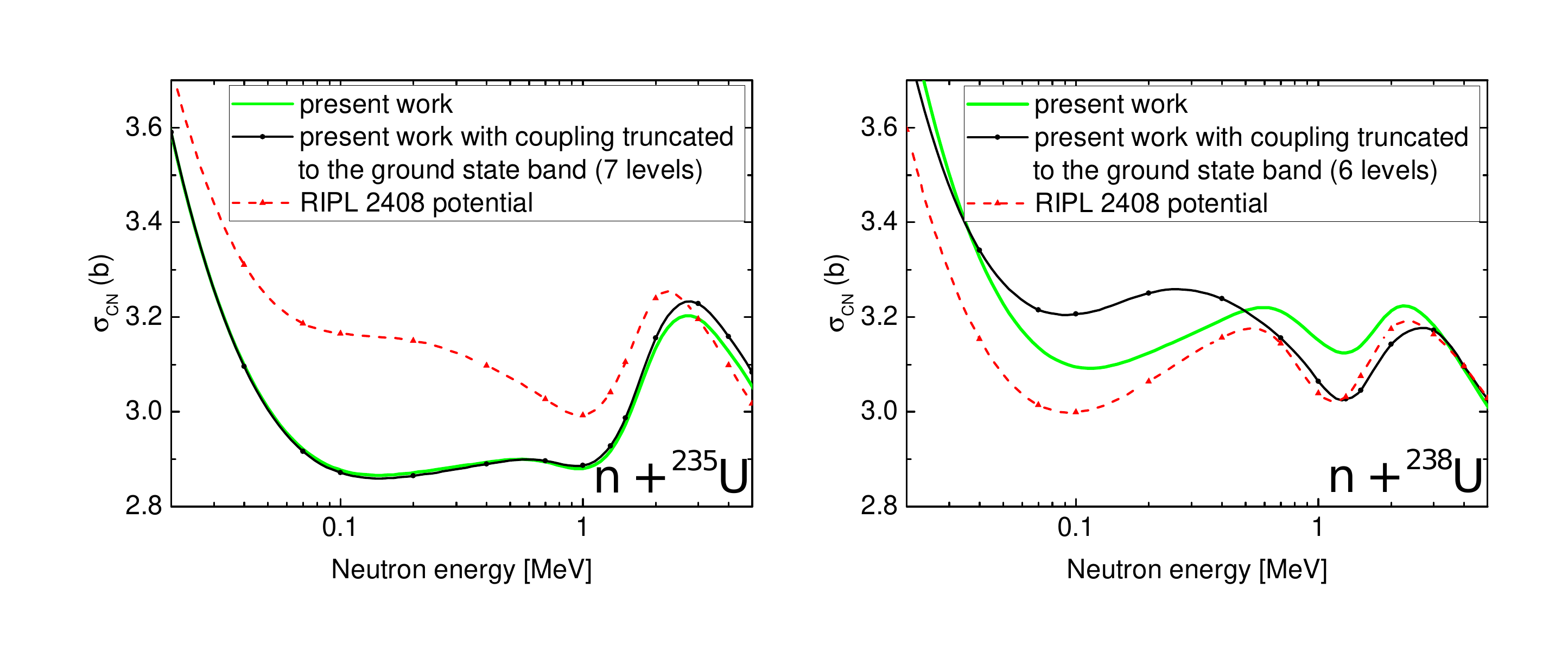}
\end{center}
\vspace{-1cm}
\caption{(Color online) Energy dependence of the calculated compound-nucleus formation cross section on $^{235}$U and $^{238}$U targets using the rigid-rotor RIPL 2408 potential \cite{casoqu05,Capote:08,ripl3} (red dashed line), and the potential derived in the current work (full green line). Calculations using the present work potential with coupling limited to the ground-state rotational band are represented by a full black line.}
\label{fig:CN}
\vspace{-0.20cm}
\end{figure*}

Dietrich \etal\cite{Dietrich:12} showed an strong impact on calculated compound-nucleus formation cross section $\sigma_{CN}(E)$, of the number of coupled levels in the rigid-rotor model. Calculated $\sigma_{CN}(E)$ for $n+^{235,238}$U reactions using rigid-rotor RIPL 2408 potential (dashed line) is compared  in Fig.~\ref{fig:CN} with calculations using the new potential with full coupling (green solid line), or with coupling reduced to GSB levels (black solid line). Left panel shows results for $^{235}$U target; calculated $\sigma_{CN}(E)$ from the potential derived in this work is much lower than the RIPL 2408 results (using 5 coupled levels) below 3~MeV. The observed difference is about 100~mb above 1~MeV ($\sim 3$\%) and reaches 300~mb ($\sim 10$\%) at 100~keV. Such differences are mainly due to the lack of saturation of the GSB coupled levels as discussed by Dietrich \etal\cite{Dietrich:12}. If we restrict the number of coupled levels for the current potential to 7 GSB levels ($K=7/2^-$ in Table VII) the results remain practically the same below 3~MeV. An small difference is observed at higher incident neutron energies due to the extended coupling for $^{235}$U target, being the full coupling result slightly lower than the one with GSB coupling. This result was expected as the single-particle coupling of excited bands in odd-$A$ nuclei is much weaker than the collective coupling for vibrational bands in even-even nuclei. For many applications, it will be a very good approximation to couple 7 levels or more of the GSB in odd-$A$ nuclei, and neglect the multiple-band coupling.

Right panel in Fig.~\ref{fig:CN} shows $\sigma_{CN}(E)$ calculations for $^{238}$U target. Again significant differences are seen below 3~MeV with the RIPL 2408 results (using 5 coupled levels) compared to present work mainly due to the lack of coupling saturation. For this even-even target, stronger coupling due to vibrational collective levels induces larger differences between the full coupling results (green solid line) and those calculated with the GSB couplings (black solid line). Such differences reach 3\%, while no differences were practically observed for odd-$A$ targets. The impact of collective levels on calculated $\sigma_{CN}(E)$ remains important from 50~keV up to 3~MeV. Therefore, we may conclude that multiple-band coupling including vibrational bands is very important for even-even targets to accurately calculate $\sigma_{CN}(E)$ in the whole energy range of interest for applications.

\section{Conclusions}

Tamura's coupling formalism~\cite{Tamura:65,Tamura:65a} has been extended to consider
low-lying bands of vibrational (single-particle) nature observed in even-even (odd-$A$) actinides.
These additional excitations are introduced as a perturbation to the underlying rigid rotor structure
that is known to describe well the ground state rotational band of major actinides.

A dispersive isospin dependent coupled-channels optical model analysis of nucleon scattering on actinides nuclei from 1~keV to 150~MeV has been undertaken. A new optical model potential is derived which couples multiple vibrational (single-particle) bands and is valid for even-even (odd-$A$) targets. The isovector terms and the very weak dependence of the geometrical parameters on mass number $A$ allowed applying the derived potential parameters to neighboring actinide nuclei with a great confidence., provided that a suitable multi-band coupling scheme could be defined. A single set of potential parameters given in Table~\ref{table:disopt} is able to describe all available scattering data on major actinides provided that the given coupling schemes and effective deformation parameters are used. While the potential is common for all actinides, the coupling scheme depends on the low-lying excitations of a given target. Fitted deformations of the ground state rotational band are in reasonable agreement with FRDM deformations theoretically derived by M\"oller and Nix \cite{Moller:95} as seen in Table \ref{table:stat-deformations}.

Excellent agreement in the whole energy range between calculations using derived dispersive coupled-channel optical model potential and the experimental total cross section for $^{242,240,239}$Pu, $^{238,235,233}$U and $^{232}$Th nuclei is obtained. A better nuclear structure description of the target nucleus also allowed an excellent description of total cross-section differences between pairs of actinide nuclei for neutron incident energies from 50~keV up to 3~MeV, fulfilling the main goal of this study: to improve the description of the neutron scattering cross sections on actinides at low neutron incident energies.

In summary, we have shown that a proposed dispersive coupled-channel phenomenological optical model with multiple-band coupling is capable of predicting \ql optical\qr nucleon induced cross section on actinides at a few percent level and gives an excellent description of the total cross-section differences among neighboring actinide nuclei (both odd and even-even) from 50~keV up to 150~MeV. Multiple-band coupling is needed for a proper calculation of compound-nucleus formation cross sections on even-even targets in the whole energy range of interest. However, it will be a very good approximation to couple only ground-state band  levels in odd-$A$ nuclei as long as the coupling is saturated (i.e., the $\sigma_{CN}(E)$ cross section does not change with increasing number of coupled levels).

Current results inspire confidence in the use of the proposed OMP with the corresponding coupling schemes to improve evaluations of neutron scattering data at low neutron incident energies as needed for many applications. Open issues like the use of the soft-rotator model to describe structure of even-even targets within the generalized optical model potential will be addressed in future works.

\section*{Acknowledgements}

This work was partially supported by International Atomic Energy Agency, through the IAEA Research Contract 13067, by the European Union and Japan under the ISTC Project B-1319, by Grants-in-Aid for Scientific Research (18560805) of JSPS and by the Spanish Ministry of Economy and Competitivity under Contracts FPA2011-28770-C03-02, FPA2014-53290-C2-2-P and FIS2011-28738-C02-01. We would like to thank anonymous referee for the extensive and detailed revision of the paper and valuable comments and suggestions which significantly improved the original manuscript. We also would like to thank our colleagues T. Kawano and Ian Thompson for their very useful advises and encouragement.

\section*{Appendix 1. Multipolar expansion of the deformed optical model
potential in the laboratory system}

$\label{Appendix1}$

Let's derive the expressions for nuclear potential expansion in spherical harmonics. If we insert nuclear shape Eq.~\eqref{radius} into the optical potential to get deformed nuclear potential, and make a Taylor expansion up
to the first order, we can obtain the following expression for the potential expansion (similar to Eq.~\eqref{pot_expansion} in section \ref{nuclear_shape})
\begin{eqnarray}
\label{eq:def-pot}
V(r,R(\theta ^{\prime },\varphi ^{\prime })) &\simeq
&V_{rot}(r,R_{axial}(\theta ^{\prime })) \\
&&+\left[ R_0 \frac{\partial }{\partial R}V(r,R(\theta ^{\prime },\varphi
^{\prime }))\right] _{R=R_{axial}(\theta ^{\prime })}\Bigg\{\beta _{20}\left[
\frac{\delta \beta _{2}}{\beta _{20}}\cos \gamma +\cos \gamma -1\right]
Y_{20}(\theta ^{\prime })  \notag \\
&&+(\beta _{20}+\delta \beta _{2})\frac{\sin \gamma }{\sqrt{2}}\left[
Y_{22}(\theta ^{\prime },\varphi ^{\prime })+Y_{2-2}(\theta ^{\prime
},\varphi ^{\prime })\right]  \notag \\
&&+\beta _{3}\left[ \cos \eta Y_{30}(\theta ^{\prime })+\frac{\sin \eta }{%
\sqrt{2}}\left[ Y_{32}(\theta ^{\prime },\varphi ^{\prime })+Y_{3-2}(\theta
^{\prime },\varphi ^{\prime })\right] \right] \Bigg\}  \notag \\
&&+\text{ (potential higher derivatives' terms)}.  \notag
\end{eqnarray}%

The expansion was made around the equilibrium axially-symmetric spheroidal
rotor shape\ $R_{axial}(\theta ^{\prime })$ (instead of the spherical shape $%
R_{0}$ used by Tamura \cite{Tamura:65}), where
\begin{equation}
R_{axial}(\theta ^{\prime })=R_{0}\left\{ 1+\underset{\lambda =2,4,6,8...}{%
\sum }\beta _{\lambda 0}Y_{\lambda 0}(\theta ^{\prime })\right\}.
\label{eq:raxial}
\end{equation}%
The radial shape $R_{axial}(\theta ^{\prime })$ corresponds to the shape
used in the traditional rigid rotor model, and is also mirror symmetric (as
only even $\lambda \geqslant 2$ are allowed in the sum). We know that the
rigid rotor model is an excellent approximation for the ground state
rotational band of actinide nuclei. On the other side, it is clear that $%
V(r,R_{0})$ and $V(r,R_{axial}(\theta ^{\prime }))$ differ. The
coupled-channel potential derived by using a Taylor expansion around the
spherical shape $R_{0}=const$ does not describe the nucleon scattering on
deformed nuclei with large deformation (e.g., rare earth, actinides, ...).
Using the rigid rotor axial potential as a zero approximation solves this
problem, therefore the aforementioned expansion is a key point in our
derivation. It should be noted that $R_{axial}(\theta ^{\prime })$=$%
R(\theta ^{\prime },\varphi ^{\prime })\vert _{(\delta \beta _{2}=0,\gamma
=0,\beta _{3}=0)}$, i.e., the axial rigid rotor corresponds to the stable
static deformation if $\delta \beta _{2}$, $\gamma $, and $\beta _{3}$
deformation parameters are equal zero.

Introducing the notation:
\begin{itemize}
\item
$V_{1}(r,\theta ^{\prime })\equiv
V_{rot}(r,R_{axial}(\theta ^{\prime }))=\left[ V(r,R(\theta ^{\prime
},\varphi ^{\prime }))\right] _{R=R_{axial}(\theta ^{\prime })}$
\item
$V_{2}(r,\theta ^{\prime })\equiv \left[  R_0 \frac{\partial }{\partial R}%
V(r,R(\theta ^{\prime },\varphi ^{\prime }))\right] _{R=R_{axial}(\theta
^{\prime })}$,
\end{itemize}
the multipole expansion of both quantities $V_{i}(r,\theta
^{\prime })$ (with $i=1,2$) in terms of the spherical harmonics with axial
symmetry $Y_{\lambda 0}(\theta ^{\prime })$ is performed:
\begin{equation}
V_{i}(r,\theta ^{\prime })=\sum_{\lambda ~(even)}v_{\lambda
}^{(i)}(r)Y_{\lambda 0}(\theta ^{\prime }),  \label{multipole1}
\end{equation}
where
\begin{equation}
v_{\lambda }^{(i)}(r)=2\pi \int_{0}^{\pi }V_{i}(r,\theta ^{\prime
})Y_{\lambda 0}(\theta ^{\prime })sin\theta ^{\prime }d\theta ^{\prime }.
\label{eq:vlambda}
\end{equation}
Note that the radial function $v_{\lambda }^{(1)}(r)$ corresponds to the
Legendre expansion coefficient of the rigid rotor potential, which is
usually already available in existing coupled channel codes.

Replacing the introduced notations into Eq.~\eqref{eq:def-pot} leads to:
\begin{eqnarray}
V(r,\theta ^{\prime },\varphi ^{\prime }) &\equiv &V(r,R(\theta ^{\prime
},\varphi ^{\prime }))=\sum_{\lambda =0,2,4,..}v_{\lambda
}^{(1)}(r)Y_{\lambda 0}(\theta ^{\prime })+\sum_{\lambda =0,2,4..}v_{\lambda
}^{(2)}(r)Y_{\lambda 0}(\theta ^{\prime })\times  \notag \\
&\Bigg \{\beta _{20}&\left[ \frac{\delta \beta _{2}}{\beta _{20}}\cos \gamma
+\cos \gamma -1\right] Y_{20}(\theta ^{\prime })  \label{potential_expanded}
\\
&+&(\beta _{20}+\delta \beta _{2})\frac{\sin \gamma }{\sqrt{2}}\left[
Y_{22}(\theta ^{\prime },\varphi ^{\prime })+Y_{2-2}(\theta ^{\prime
},\varphi ^{\prime })\right]  \notag \\
&+&\beta _{3}\left[ \cos \eta Y_{30}(\theta ^{\prime })+\frac{\sin \eta }{%
\sqrt{2}}\left[ Y_{32}(\theta ^{\prime },\varphi ^{\prime })+Y_{3-2}(\theta
^{\prime },\varphi ^{\prime })\right] \right] \Bigg \} . \notag
\end{eqnarray}

By using the formula for the product of two spherical harmonics and
replacing $Y_{\lambda \nu }(\theta ^{\prime },\varphi ^{\prime })$
(expressed in the intrinsic system of reference) by $\sum_{\mu }D_{\mu ,\nu
}^{\lambda }Y_{\lambda \mu }(\theta ,\varphi )$ where $\theta $ and $\varphi
$ are the polar angles referred to the space-fixed coordinates and $D_{\mu
,\nu }^{\lambda }$ are the rotation Wigner functions, one obtains:
\begin{eqnarray}
V(r,\theta ,\varphi )&&=\sum_{\lambda =0,2,4,..}v_{\lambda
}^{(1)}(r)\sum_{\mu }D_{\mu ,0}^{\lambda }Y_{\lambda \mu }(\theta ,\varphi )
\label{V_multip_exp} \\
&&+\beta _{20}\left[ \frac{\delta \beta _{2}}{\beta _{20}}\cos \gamma +\cos
\gamma -1\right] \sum_{\lambda =0,2,4,..}\left[ \tilde{v}_{\lambda }^{(2)}(r)%
\right] _{0}\frac{1}{\sqrt{2}}\sum_{\mu }D_{\mu ,0}^{\lambda }Y_{\lambda \mu
}(\theta ,\varphi )  \notag \\
&&+(\beta _{20}+\delta \beta _{2})\frac{\sin \gamma }{\sqrt{2}}\sum_{\lambda
=2,4,6,..}\left[ \tilde{v}_{\lambda }^{(2)}(r)\right] _{2}\sum_{\mu }\left[
(D_{\mu ,2}^{\lambda }+D_{\mu ,-2}^{\lambda })Y_{\lambda \mu }(\theta
,\varphi )\right]  \notag \\
&&+\beta _{3}\left[ \cos \eta \sum_{\lambda =1,3,5,..}\left[ \tilde{v}%
_{\lambda }^{(3)}(r)\right] _{0}\sum_{\mu }D_{\mu ,0}^{\lambda }Y_{\lambda
\mu }(\theta ,\varphi )\right] \notag \\
&&+\beta _{3}\left[ \frac{\sin \eta }{\sqrt{2}}\sum_{\lambda =3,5,7,..}%
\left[ \tilde{v}_{\lambda }^{(3)}(r)\right] _{2}\sum_{\mu }\left[ (D_{\mu
,2}^{\lambda }+D_{\mu ,-2}^{\lambda })Y_{\lambda \mu }(\theta ,\varphi )%
\right] \right]  \notag \\
&&=V_{diag}(r)+V_{coupling}(r,\theta ,\varphi ),  \notag
\end{eqnarray}%
where $V_{diag}$ is the $\lambda =\mu =0$ component of the first term of Eq.~%
\eqref{V_multip_exp}, while $V_{coupling}$ is the rest and :\newline
\begin{eqnarray}
\left[ \tilde{v}_{\lambda }^{(2)}(r)\right] _{0} &=&\sum_{\lambda ^{\prime
}=0,2,4,..}v_{\lambda ^{\prime }}^{(2)}(r)\left[ \frac{5(2\lambda ^{\prime
}+1)}{4\pi (2\lambda +1)}\right] ^{1/2}\left\langle \lambda ^{\prime }200\vert\lambda 0\right\rangle ^{2},
\label{v20} \\
\left[ \tilde{v}_{\lambda }^{(2)}(r)\right] _{2} &=&\sum_{\lambda ^{\prime
}=0,2,4,...}v_{\lambda ^{\prime }}^{(2)}(r)\left[ \frac{5(2\lambda ^{\prime
}+1)}{4\pi (2\lambda +1)}\right] ^{1/2}\left\langle \lambda ^{\prime }200\vert\lambda
0\right\rangle \left\langle \lambda ^{\prime }202\vert\lambda 2\right\rangle ,  \label{v22} \\
\left[ \tilde{v}_{\lambda }^{(3)}(r)\right] _{0} &=&\sum_{\lambda ^{\prime
}=0,2,4,..}v_{\lambda ^{\prime }}^{(2)}(r)\left[ \frac{7(2\lambda ^{\prime
}+1)}{4\pi (2\lambda +1)}\right] ^{1/2}\left\langle \lambda ^{\prime }300\vert\lambda 0\right\rangle ^{2},
\label{v30} \\
\left[ \tilde{v}_{\lambda }^{(3)}(r)\right] _{2} &=&\sum_{\lambda ^{\prime
}=0,2,4,..}v_{\lambda ^{\prime }}^{(2)}(r)\left[ \frac{7(2\lambda ^{\prime
}+1)}{4\pi (2\lambda +1)}\right] ^{1/2}\left\langle \lambda ^{\prime }300\vert\lambda
0\right\rangle \left\langle \lambda ^{\prime }302\vert\lambda 2\right\rangle ,  \label{v32}
\end{eqnarray}%
with the following constraints:

\begin{itemize}
\item[-] Only {\bfseries even $\lambda ^{\prime }$} values are allowed in
the summations which appear in the expressions of $\tilde{v}_{\lambda
}^{(2)}(r)$ and $\tilde{v}_{\lambda }^{(3)}(r)$ in Eqs.~\eqref{v20} to %
\eqref{v32}, since $v_{\lambda ^{\prime }}^{(2)}=0$ for odd $\lambda
^{\prime }$ (from the axial and reflection symmetry of the equilibrium
nuclear shape).

\item[-] Only {\bfseries even $\lambda $} values are allowed for the
quadrupole vibrational coupling $\tilde{v}_{\lambda }^{(2)}(r)$ because of
the\ $\left\langle \lambda ^{\prime }200\vert\lambda 0\right\rangle $ Clebsch-Gordan coefficient in Eqs.~%
\eqref{v20} and \eqref{v22}, since $\lambda ^{\prime }$ is even.

\item[-] Only {\bfseries odd $\lambda $} values are allowed for the octupole
vibrational coupling term $\tilde{v}_{\lambda }^{(3)}(r)$ because of the $%
\left\langle \lambda ^{\prime }300\vert\lambda 0\right\rangle $ Clebsch-Gordan coefficient in Eqs.~%
\eqref{v30} and \eqref{v32}, since $\lambda ^{\prime }$ is even.

\item[-] $[\tilde{v}_{0}^{(2)}]_{2}=[\tilde{v}_{1}^{(3)}]_{2}=0$ due to the
second Clebsch-Gordan coefficient in Eqs.~\eqref{v22} and \eqref{v32}, as
projection $\mu =2$ can not be larger than corresponding momentum $\lambda $.
\end{itemize}

The deformed optical model potential of Eq.~\eqref{V_multip_exp}, through
its variables $\delta \beta _{2}$, $\gamma $, $\beta _{3}$, and $\eta$
($\beta _{20}$ is the static deformation and doesn't play any role in the
couplings) guarantees coupling of vibrational bands (coupling between
rotational states, described by the Wigner's $D$-funtions, are made
through the spherical harmonics). Thus, for example, the first term usually corresponds
to the intra-band coupling (e.g., couples the rotational band members built
on the same vibrational state); the second term couples the ground-state band with quadrupolar vibrational
states (e.g., $\beta $-vibrational and $\gamma $-vibrational bands); the third term couples states
in the ground-state band with states of the octupolar vibrational band;  and so on.

\section*{Appendix 2. Coupled-channel matrix elements}

{\label{Appendix2} }

We will show how to calculate coupling matrix elements describing transitions between nuclear levels $\vert i\rangle $ and $\vert f\rangle $ due to the deformed potential (using potential expansion \eqref{V_multip_exp} derived in Appendix 1), assuming that the incident projectile is a nucleon with spin $s=1/2$ . Following Tamura's notation and procedures from Refs.~\cite{Tamura:65,Tamura:65a}, and using the nuclear wave functions, we can calculate the coupled-channel matrix elements needed in reaction calculations. As discussed we have to analyze the even-even and odd-nuclei separately.

\subsection{Even-even nuclei}

The nuclear wavefunction $\Psi \equiv \vert i\rangle $ was defined in Eq.~\eqref{wave-even}. 
Using the even-even wave function for both the initial and final nuclear states $\vert i\rangle$ and $\vert f\rangle$
(where $\overrightarrow{j}=\overrightarrow{l}+\overrightarrow{s}$ is the total angular momentum of the projectile equal to the vector sum of the orbital angular momentum $\overrightarrow{l}$ and the projectile spin $\overrightarrow{s}=\frac12 $), and following Eqs.~(26-29) given by Tamura \cite{Tamura:65}, we can derive the coupled-channel matrix elements in terms of the reduced matrix elements as follows,
\begin{eqnarray}
\left\langle i\vert V(r,\theta ,\varphi )\vert f \right\rangle &&=\sum_{K}^{I}~A_{K}^{I\tau }\sum_{K^{\prime
}}^{I^{\prime }}~A_{K^{\prime }}^{I^{\prime }\tau ^{\prime }}~\Bigg\{\sum_{\lambda
=0,2,4..}v_{\lambda }^{(1)}(r) \left\langle IK \vert\vert D_{;0}^{\lambda } \vert\vert I^{\prime}K\right\rangle
A(ljI;j^{\prime }l^{\prime }I^{\prime };\lambda J\frac12)~\delta _{K,K^{\prime }} \notag \\
&& +\left\langle n_i(\lambda=2) \right\vert \beta _{20}\left[ \frac{\delta \beta _{2}}{\beta _{20}}\cos \gamma +\cos
\gamma -1\right] \left\vert n_f(\lambda=2)\right\rangle  \notag \\
&& \:\:\:\:\:\: \times \sum_{\lambda =0,2,4,..}[\tilde{v}_{\lambda}^{(2)}(r)]_{0}\left\langle IK\vert\vert D_{;0}^{\lambda }\vert\vert I^{\prime }K\right\rangle
A(ljI;j^{\prime}l^{\prime }I^{\prime };\lambda J\frac12 )~\delta _{K,K^{\prime }}  \notag \\
&& + \left\langle n_i(\lambda=2) \right\vert (\beta _{20}+\delta \beta _{2})\frac{\sin \gamma }{\sqrt{2}}%
\left\vert n_f(\lambda=2)\right\rangle \notag \\
&& \:\:\:\:\:\: \times \sum_{\lambda =2,4,6,..}[\tilde{v}_{\lambda }^{(2)}(r)]_{2}\left\langle IK\vert\vert \left(
D_{;2}^{\lambda }+D_{;-2}^{\lambda }\right) \vert\vert I^{\prime }K^{\prime}\right\rangle
A(ljI;j^{\prime }l^{\prime }I^{\prime };\lambda J{\frac12})  \label{coupling_jmq} \\
&&+ \left\langle n_i(\lambda=3)\right\vert \beta _{3}\cos \eta \left\vert n_f(\lambda=3)\right\rangle \notag \\
&& \:\:\:\:\:\: \times \sum_{\lambda =1,3,5,..}[\tilde{v}_{\lambda
}^{(3)}(r)]_{0}\left\langle IK\vert\vert D_{;0}^{\lambda }\vert\vert I^{\prime }K\right\rangle A(ljI;j^{\prime
}l^{\prime }I^{\prime };\lambda J{\frac12})~\delta _{K,K^{\prime }}  \notag \\
&&+ \left\langle n_i(\lambda=3) \right\vert \beta _{3}\frac{\sin \eta }{\sqrt{2}} \left\vert n_f(\lambda=3)\right\rangle \notag \\
&& \:\:\:\:\:\: \times \sum_{\lambda =3,5,7,..}[\tilde{v}_{\lambda }^{(3)}(r)]_{2}\left\langle IK\vert\vert \left( D_{;2}^{\lambda }+D_{;-2}^{\lambda
}\right) \vert\vert I^{\prime }K^{\prime }\right\rangle A(ljI;j^{\prime }l^{\prime }I^{\prime};\lambda J{\frac12})\Bigg\}.  \notag
\end{eqnarray}%
Note that the collective variables $\delta\beta_2$, $\gamma$, $\beta_3$, and $\eta$ are averaged
over the corresponding initial $\left\vert n_i(\lambda) \right\rangle$ and final $\left\vert n_f(\lambda)\right\rangle$
vibrational wave functions. The diagonal elements in $K$ and $K^{\prime }$ correspond to the case $\mu
=0 $ ($\Delta K=0$ transition), and usually describe the intra-band
transitions. If we neglect 
small deformation parameters ($\delta \beta _{2}
=\beta _{3}=0$), and assume axial symmetry ($\gamma =0$), then the coupling matrix element Eq.~\eqref{coupling_jmq} is simplified to the one equivalent to the axial rigid rotor,

\begin{equation}
\left\langle i\vert V(r,\theta ,\varphi )\vert f\right\rangle =\sum_{\lambda =0,2,4..}v_{\lambda
}^{(1)}(r)\left\langle IK\vert\vert D_{;0}^{\lambda }\vert\vert I^{\prime }K\right\rangle A(ljI;j^{\prime }l^{\prime
}I^{\prime };\lambda J%
{\frac12}%
)~\delta _{K,K^{\prime }}\;. \label{coupling_axial_rigid_rotor}
\end{equation}%
The pure geometrical factor $A(ljI;j^{\prime }l^{\prime }I^{\prime };\lambda J{\frac12})$
was given by Tamura (see Eq.~(28) in Ref.~\cite{Tamura:65} written for $%
\overrightarrow{s}$=${\frac12}$. Note a typo, there is a missing hat on $l$ and $l^{\prime }$).
The $A$ geometrical factor depends on the total momentum of the system $%
J $ and is given by,
\begin{eqnarray}
\label{A}
A(ljI;l^{\prime }j^{\prime }I^{\prime };\lambda J{\frac12})&=&
\frac{1}{4\pi }(-1)^{J-\frac{1}{2}-I^{\prime }+l+l^{\prime }+\frac{1}{2}%
(l^{\prime }-l)}\sqrt{(2l+1)(2l^{\prime }+1)(2j+1)(2j^{\prime }+1)}\left\langle ll^{\prime
}00\vert\lambda 0\right\rangle  \times \notag \\
&& W(jIj^{\prime }I^{\prime };J\lambda )~W(ljl^{\prime }j^{\prime};
{\frac12}\lambda ),
\end{eqnarray}
where $\left\langle ll^{\prime }00\vert\lambda 0\right\rangle $ corresponds to the
Clesbch-Gordan and $W(jIj^{\prime }I^{\prime };J\lambda )$ to the Racah
coefficients \cite{Tamura:65,Tamura:65a}. It is important to remark that the
coefficient $A(ljI;l^{\prime }j^{\prime }I^{\prime };\lambda J{\frac12})$
is present in all sums in Eq.~\eqref{coupling_jmq}, and contains the
Clebsch-Gordan coefficient $\left\langle ll^{\prime }00\vert\lambda 0\right\rangle $ that imposes the
parity selection rule on the orbital momentum $l$,$l^{\prime }$ of the
incident nucleon and the transition multipolarity $\lambda $ that
$l+l^{\prime }+\lambda $ has to be an even number.

The expression of the coupling potential given by Eq.~\eqref{coupling_jmq}
is general under the assumption of small deformations around equilibrium
axially symmetric shape (i.e. $\delta \beta _{2}$, $\gamma $, and $\beta _{3}$ are small). If we further restrict to the monopole component of the axial derivative of the potential $V_{2}(r,\theta ^{\prime })$ since it carries
the largest contribution, the following expression follows:
\begin{eqnarray}
\left\langle i\vert V(r,\theta ,\varphi )\vert f\right\rangle  &&=\sum_{K}^{I}~A_{K}^{I\tau }\sum_{K^{\prime
}}^{I^{\prime }}~A_{K^{\prime }}^{I^{\prime }\tau ^{\prime }}~\Bigg\{\sum_{\lambda
=0,2,4,..}v_{\lambda }^{(1)}(r)\left\langle IK\vert\vert D_{;0}^{\lambda }\vert\vert I^{\prime
}K\right\rangle A(ljI;l^{\prime }j^{\prime }I^{\prime };\lambda J%
{\frac12}%
)~\delta _{K,K^{\prime }}+\frac{v_{0}^{(2)}(r)}{\sqrt{4\pi }}\Bigg\{  \notag
\\
&&\left\langle n_i(\lambda=2) \right\vert \beta _{20}\left[ \frac{\delta \beta _{2}}{\beta _{20}}\cos \gamma +\cos
\gamma -1\right] \left\vert n_f(\lambda=2)\right\rangle ~\left\langle IK\vert\vert D_{;0}^{2}\vert\vert I^{\prime }K\right\rangle A(ljI;l^{\prime }j^{\prime
}I^{\prime };2J%
{\frac12}%
)~\delta _{K,K^{\prime }}+  \notag \\
&&\left\langle n_i(\lambda=2) \right\vert (\beta _{20}+\delta \beta _{2})\frac{\sin \gamma }{\sqrt{2}} \left\vert n_f(\lambda=2)\right\rangle ~\left\langle IK\vert\vert \left(
D_{;2}^{2}+D_{;-2}^{2}\right) \vert\vert I^{\prime }K^{\prime }\right\rangle A(ljI;l^{\prime
}j^{\prime }I^{\prime };2J%
{\frac12}%
)+  \label{matrix-elem-monop-approx} \\
&&\left\langle n_i(\lambda=3) \right\vert \beta _{3}\cos \eta \left\vert n_f(\lambda=3)\right\rangle ~\left\langle IK\vert\vert D_{;0}^{3}\vert\vert I^{\prime }K\right\rangle A(ljI;l^{\prime
}j^{\prime }I^{\prime };3J%
{\frac12}%
)~\delta _{K,K^{\prime }}+  \notag \\
&&\left\langle n_i(\lambda=3) \right\vert \beta _{3}\frac{\sin \eta }{\sqrt{2}}\left\vert n_f(\lambda=3)\right\rangle ~\left\langle IK\vert\vert \left(
D_{;2}^{3}+D_{;-2}^{3}\right) \vert\vert I^{\prime }K^{\prime }\right\rangle A(ljI;l^{\prime
}j^{\prime }I^{\prime };3J%
{\frac12}%
)\Bigg\}\Bigg\}.  \notag
\end{eqnarray}%
To derive the expression \eqref{matrix-elem-monop-approx} above, we have
considered the fact that the terms which contribute with the main
(monopolar) $v_{0}^{(2)}(r)$ component of $V_{2}(r,\theta ^{\prime })$ are
the ones with $\lambda =2$ for quadrupolar and $\lambda =3$ for octupolar
deformations, which automatically leads to keep in Eq.~\eqref{coupling_jmq}
only the terms%
\begin{equation}
\lbrack \tilde{v}_{2}^{(2)}(r)]_{0}=[\tilde{v}_{2}^{(2)}(r)]_{2}=[\tilde{v}%
_{3}^{(3)}(r)]_{0}=[\tilde{v}_{3}^{(3)}(r)]_{2}\simeq \frac{v_{0}^{(2)}(r)}{%
\sqrt{4\pi }}.
\end{equation}

Moreover, the monopole component $v_{0}^{(2)}(r)$ of $V_{2}(r,\theta
^{\prime })$ can be shown to be proportional to the quadrupole component of
the axial potential $V_{rot}(r,R_{axial}(\theta ^{\prime }))$, if we restrict
to the first order term of its expansion in the parameter $\beta _{20}$

\begin{eqnarray}
V_{rot}(r,R_{axial}(\theta ^{\prime })) &=&v_{0}(r)+\left[ R_0 \frac{\partial }{\partial R_{axial}}V_{rot}(r,R_{axial}(\theta ^{\prime }))\right] _{\beta
_{\lambda 0}=0}\times \underset{\lambda =2,4,6,..}{\sum }\beta _{\lambda
0}Y_{\lambda 0}(\theta ^{\prime })+  \notag \\
&&\sum\limits_{t>1}\left[ R^t_0 \frac{\partial ^{t}}{\partial ^{t}R_{axial}}%
V_{rot}(r,R_{axial}(\theta ^{\prime }))\right] _{\beta _{\lambda 0}=0}\times %
\left( \underset{\lambda =2,4,6,..}{\sum }\beta _{\lambda 0}Y_{\lambda
0}(\theta ^{\prime })\right)^{t}\;.  \label{potexpand}
\end{eqnarray}

Performing the multipole expansion of the axial rigid rotor potential in
spherical harmonics $Y_{\lambda 0}(\theta ^{\prime })$ according to Eq.~%
\eqref{multipole1} and comparing with Eq.~\eqref{potexpand}, we obtain the
expressions for expansion coefficients $v_{0}^{(1)}(r)$ and $v_{0}^{(2)}(r)$ as follows,
\begin{eqnarray}
v_{0}^{(1)}(r)Y_{00} &=&v_{0}(r)+\sum\limits_{t>1}\left[ R^t_0 \frac{\partial ^{t}}{\partial ^{t}R_{axial}}V_{rot}(r,R_{axial}(\theta ^{\prime }))\right]
_{\beta _{\lambda 0}=0}\times \left[ \underset{\lambda =2,4,6...}{\sum \beta
_{\lambda 0}}Y_{\lambda 0}(\theta ^{\prime }) \right] _{00}^{t}  \notag \\
&=&v_{0}(r)+O(\beta _{\lambda 0}^{2})\simeq v_{0}(r),  \label{key}
\end{eqnarray}

\begin{eqnarray}
v_{2}^{(1)}(r)Y_{20}(\theta ^{\prime }) &=&\left[ R_0\frac{\partial }{\partial
R_{axial}}V_{rot}(r,R_{axial}(\theta ^{\prime }))\right] _{\beta_{\lambda 0}=0}\beta _{20}Y_{20}(\theta ^{\prime }) \\
&&+\sum\limits_{t>1}\left[ R^t_0 \frac{\partial ^{t}}{\partial ^{t}R_{axial}}%
V_{rot}(r,R_{axial}(\theta ^{\prime }))\right] _{\beta _{\lambda 0}=0}\times \left[
\underset{\lambda =2,4,6...}{\sum \beta _{\lambda 0}}Y_{\lambda
0}(\theta ^{\prime }) 
\right] _{20}^{t}  \notag \\
&=&\left[ R_0\frac{\partial }{\partial R_{axial}}V_{rot}(r,R_{axial}(\theta
^{\prime }))\right] _{\beta _{\lambda 0}=0}\beta _{20}Y_{20}(\theta ^{\prime
})+O(\beta _{\lambda 0}^{2})  \notag \\
&\simeq &\left[R_0 \frac{\partial }{\partial R_{axial}}V_{rot}(r,R_{axial}(%
\theta ^{\prime }))\right] _{\beta _{\lambda 0}=0}\beta _{20}Y_{20}(\theta ^{\prime
})\;.  \notag
\end{eqnarray}%

And finally we obtain:
\begin{equation}
\left[ R_0\frac{\partial }{\partial R_{axial}}V_{rot}(r,R_{axial}(\theta
^{\prime }))\right] _{\beta _{\lambda 0}=0}\simeq \frac{v_{2}^{(1)}(r)}{\beta _{20}}\;,
\end{equation}%
which on the other hand is the monopole term of the axially simmetric
derivative of the potential, $V_{2}(r,\theta ^{\prime })$. Therefore:

\begin{equation}
\frac{v^{(1)}_{2}(r)}{\beta _{20}} \simeq \frac{v_0^{(2)}(r)}{\sqrt{ 4 \pi}}\;.
\label{approx}
\end{equation}

Including Eq.~\eqref{approx} into the Eq.~\eqref{matrix-elem-monop-approx},
we get the expression
\begin{eqnarray}
\left\langle i\vert V(r,\theta ,\varphi )\vert f\right\rangle  &&=\sum_{K}^{I}\sum_{K^{\prime }}^{I^{\prime
}}A_{K}^{I\tau }A_{K^{\prime }}^{I^{\prime }\tau ^{\prime }}~\Bigg\{\sum_{\lambda
=0,2,4,..}v_{\lambda }^{(1)}(r)\left\langle IK\vert\vert D_{;0}^{\lambda }\vert\vert I^{\prime
}K\right\rangle A(ljI;l^{\prime }j^{\prime }I^{\prime };\lambda J%
{\frac12}%
)~\delta _{K,K^{\prime }}+\frac{v_{2}^{(1)}(r)}{\beta _{20}}\times \Bigg\{
\notag \\
&& \left\langle n_i(\lambda=2) \right\vert \beta _{20}\left[ \frac{\delta \beta _{2}}{\beta _{20}}\cos \gamma +\cos
\gamma -1\right] \left\vert n_f(\lambda=2)\right\rangle ~\left\langle IK\vert\vert D_{;0}^{2}\vert\vert I^{\prime }K\right\rangle A(ljI;l^{\prime }j^{\prime}I^{\prime };2J%
{\frac12}%
)~\delta _{K,K^{\prime }}~+  \notag \\
&& \left\langle n_i(\lambda=2) \right\vert (\beta _{20}+\delta \beta _{2})\frac{\sin \gamma }{\sqrt{2}} \left\vert n_f(\lambda=2)\right\rangle ~\left\langle IK\vert\vert \left(D_{;2}^{2}+D_{;-2}^{2}\right) \vert\vert I^{\prime }\pi ^{\prime }K^{\prime}\right\rangle A(ljI;l^{\prime }j^{\prime }I^{\prime };2J%
{\frac12}%
)~+ \notag \\
&&\left\langle n_i(\lambda=3) \right\vert \beta _{3}\cos \eta \left\vert n_f(\lambda=3)\right\rangle ~\left\langle I\pi K\vert\vert D_{;0}^{3}\vert\vert I^{\prime }\pi ^{\prime}K\right\rangle A(ljI;l^{\prime }j^{\prime }I^{\prime };3J%
{\frac12}%
)~\delta _{K,K^{\prime }}~+  \label{matrix-element-no-aver} \\
&&\left\langle n_i(\lambda=3) \right\vert \beta _{3}\frac{\sin \eta }{\sqrt{2}}\left\vert n_f(\lambda=3)\right\rangle ~\left\langle I\pi K\vert\vert \left(D_{;2}^{3}+D_{;-2}^{3}\right) \vert\vert I^{\prime }\pi ^{\prime }K^{\prime}\right\rangle A(ljI;l^{\prime }j^{\prime }I^{\prime };3J{\frac12}
)\Bigg\}\Bigg\}\;,  \notag
\end{eqnarray}%
where all radial functions $v_{\lambda }^{(1)}(r)$ corresponds to the
Legendre expansion coefficient of the rigid rotor potential.

To the lowest order in the dynamic deformations (i.e., all products of dynamical deformation variables
are neglected), which are averaged over the corresponding vibrational
wave functions, the expression which has been coded in OPTMAN is obtained:
\begin{eqnarray}
\left\langle i\vert V(r,\theta ,\varphi )\vert f\right\rangle  &&=\sum_{K}^{I}\sum_{K^{\prime }}^{I^{\prime
}}A_{K}^{I\tau }A_{K^{\prime }}^{I^{\prime }\tau ^{\prime }}~\Bigg\{\sum_{\lambda
=0,2,4,..}v_{\lambda }^{(1)}(r)\left\langle IK\vert\vert D_{;0}^{\lambda }\vert\vert I^{\prime
}K\right\rangle A(ljI;l^{\prime }j^{\prime }I^{\prime };\lambda J{\frac12})~\delta _{K,K^{\prime }}+v_{2}^{(1)}(r)\times \Bigg\{  \notag \\
&&\left[ \left[ \beta _{2}\right] _{eff}+\left[ \gamma _{20}\right] _{eff}%
\right] ~\left\langle IK\vert\vert D_{;0}^{2}\vert\vert I^{\prime }K\right\rangle A(ljI;l^{\prime }j^{\prime }I^{\prime
};2J{\frac12})~\delta _{K,K^{\prime }}~+  \notag \\
&&\left[ \gamma _{22}\right] _{eff}~\left\langle IK\vert\vert \left(
D_{;2}^{2}+D_{;-2}^{2}\right) \vert\vert I^{\prime }\pi ^{\prime }K^{\prime
}\right\rangle A(ljI;l^{\prime }j^{\prime }I^{\prime };2J{\frac12})~+  \label{matrix-element} \\
&&\left[ \beta _{30}\right] _{eff}~\left\langle I\pi K\vert\vert D_{;0}^{3}\vert\vert I^{\prime }\pi
^{\prime }K\right\rangle A(ljI;l^{\prime }j^{\prime }I^{\prime };3J{\frac12})~\delta _{K,K^{\prime }}~+  \notag \\
&&\left[ \beta _{32}\right] _{eff}~\left\langle I\pi K\vert\vert \left(
D_{;2}^{3}+D_{;-2}^{3}\right) \vert\vert I^{\prime }\pi ^{\prime }K^{\prime
}\right\rangle A(ljI;l^{\prime }j^{\prime }I^{\prime };3J%
{\frac12})\Bigg\}\Bigg\}\;,  \notag
\end{eqnarray}%
where the radial functions $v_{\lambda }^{(1)}(r)$ are calculated by Eq.~\eqref{eq:vlambda}.

Considering the orthogonality of the vibrational wave functions $\left\langle n_i(\lambda)\vert n_f(\lambda)\right\rangle =\delta_{if}$,
the effective deformation coefficients defining the inter-band coupling strength are given as follow
\begin{itemize}
\item $\left[ \beta_{2}\right] _{eff}\equiv \left\langle n_{i}(\lambda=2) \right\vert \frac{\delta \beta_{2}}{\beta _{20}} \left\vert n_{f}(\lambda=2)\right\rangle =
\left\langle n_{i}(\beta) \right\vert \frac{\delta \beta_{2}}{\beta _{20}} \left\vert n_{f}(\beta)\right\rangle $,

\item $\left[ \gamma_{20}\right] _{eff}\equiv \left\langle n_i(\lambda=2) \right\vert \left[\cos \gamma -1\right] \left\vert n_f(\lambda=2)\right\rangle =
       \left\langle n_{i}(\gamma) \right\vert \left[\cos \gamma -1\right] \left\vert n_{f}(\gamma)\right\rangle
    \approx -\frac12 \left\langle n_{i}(\gamma) \right\vert \left[\gamma ^{2}\right] \left\vert n_{f}(\gamma)\right\rangle$,

\item $\left[ \gamma _{22}\right]_{eff}\equiv
\left\langle n_i(\lambda=2) \right\vert \left[ \frac{1}{\sqrt{2}}\sin \gamma \right] \left\vert n_f(\lambda=2)\right\rangle =
\left\langle n_i(\gamma) \right\vert \left[ \frac{1}{\sqrt{2}}\sin \gamma \right] \left\vert n_f(\gamma)\right\rangle \approx
\frac{1}{\sqrt{2}} \left\langle n_{i}(\gamma) \right\vert \left[ \gamma \right] \left\vert n_{f}(\gamma)\right\rangle$,

\item $\left[ \beta _{30}\right]_{eff}\equiv
\left\langle n_{i}(\lambda=3) \right\vert \left[ \frac{\beta _{3}}{\beta _{20}}\cos \eta \right] \left\vert n_{f}(\lambda=3)\right\rangle$, and

\item $\left[ \beta _{32}\right] _{eff}\equiv
\left\langle n_{i}(\lambda=3) \right\vert \left[ \frac{1}{\sqrt{2}}\frac{\beta _{3}}{\beta _{20}}\sin \eta \right] \left\vert n_{f}(\lambda=3) \right\rangle$.
\end{itemize}

Note that the $\cos \gamma $ Taylor expansion should be taken up to the second order
to consider axial transitions between the excited quadrupolar $\gamma $-band
($K=0^{+}$) and the ground state ($K=0^{+}$), otherwise no such transitions will be
allowed. The above defined deformation factors can be treated as \ql effective\qr
coupling strength of the excited bands as follow,

\begin{itemize}
\item $\left[ \beta _{2}\right] _{eff}-$ quadrupolar ($\lambda =2$) axial
transitions with $\Delta K=\mu =0$ (e.g., between the $K=0^+$~$\beta$-band
and the ground state, or intra-band transitions);

\item $\left[ \gamma _{20}\right] _{eff}-$ quadrupolar ($\lambda =2$) axial
transitions with $\Delta K=\mu =0$ (e.g., between the $K=0^+$~$\gamma$-band
and the ground state);

\item $\left[ \gamma _{22}\right] _{eff}-$ quadrupolar ($\lambda =2$)
non-axial transitions with $\Delta K=\mu =\pm 2$ (e.g., between the $K=2^+$
non-axial quadrupolar band and the ground state);

\item $\left[ \beta _{30}\right] _{eff}-$ octupolar ($\lambda =3$) axial
transitions with $\Delta K=\mu =0$ (e.g., between the $K=0^-$ axial
octupolar band and the ground state), and

\item $\left[ \beta _{32}\right] _{eff}-$ octupolar ($\lambda =3$) non-axial
transitions with $\Delta K=\mu =\pm 2$ (e.g., between the $K=2^-$ non-axial
octupolar band and the ground state).
\end{itemize}

If we had allowed for octupolar vibrations $\lambda=3$ with odd projections ($\mu=\pm 1,\pm 3$),
then we would have two additional effective deformations $\left[ \beta _{31}\right] _{eff}$ and
$\left[ \beta _{33}\right] _{eff}$ that correspond to octupolar ($\lambda =3$) non-axial
transitions with $\Delta K =1$ ($\mu=\pm 1$) and $\Delta K=3$ ($\mu=\pm3$). Those transitions
add an additional complexity to the coupling scheme if they are taken into account.

Considering the corresponding wave functions, the matrix elements in Eq.~%
\eqref{matrix-element} above contain the product of three Wigner functions
which can be calculated by the expression given by Davidov \cite{davidov62},
\begin{equation}
\int_{\Omega }^{{}}D_{MK}^{\ast J}(\Omega )D_{m_{1}k_{1}}^{j_{1}}(\Omega
)D_{m_{2}k_{2}}^{j_{2}}(\Omega )d\Omega =\frac{8\pi ^{2}}{2J+1}%
\left\langle j_{1}j_{2}m_{1}m_{2}\vert JM\right\rangle \left\langle j_{1}j_{2}k_{1}k_{2}\vert JK\right\rangle \;.
\end{equation}%
The general expression for coupling reduced matrix elements in even-even nucleus is
given below,
\begin{eqnarray}
\left\langle IK\vert\vert \left[ D_{;\mu }^{\lambda }+(-1)^{\mu }D_{;-\mu }^{\lambda }\right]
\vert\vert I^{\prime }K^{\prime }\right\rangle  &=&\frac{\sqrt{2I^{\prime }+1}}{\sqrt{(1+\delta
_{K0})(1+\delta _{K^{\prime }0})}} \; \frac{\lbrack 1+(-1)^{\lambda +\mu
+\lambda _{ph}+\lambda _{ph}^{\prime }}]}{2}\times  \label{red-mat-even-gen}
\\
&&\big[\left\langle I^{\prime }\lambda K^{\prime }\mu \vert IK\right\rangle +~(-1)^{I^{\prime }+\lambda
_{ph}^{\prime }}\left\langle I^{\prime }\lambda -K^{\prime }\mu \vert IK\right\rangle +  \notag \\
&&(-1)^{I+\lambda _{ph}}\left\langle I^{\prime }\lambda K^{\prime }\mu
\vert I-K\right\rangle +~(-1)^{I+I^{\prime }+\lambda _{ph}+\lambda _{ph}^{\prime }}\left\langle I^{\prime
}\lambda -K^{\prime }\mu \vert I-K\right\rangle \big]\;,  \notag
\end{eqnarray}%
where $(-1)^{\lambda +\mu +\lambda _{ph}+\lambda _{ph}^{\prime }}=(-1)^{\lambda +\lambda
_{ph}+\lambda _{ph}^{\prime }}$ (as $\mu =even$) defining the following
selection rules:

- if the transition occurs intra-band then $\lambda _{ph}=\lambda
_{ph}^{\prime }$, therefore $\lambda =even$;

- if the transition occurs between quadrupolar vibrational bands and the
ground state then $\lambda _{ph}=0,\lambda _{ph}^{\prime }=2$ (or $\lambda
_{ph}=2,\lambda _{ph}^{\prime }=0$), therefore $\lambda =even$ (only
quadrupolar transitions allowed);

- if the transition occurs between octupolar vibrational bands and the
ground state $\lambda _{ph}=0,\lambda _{ph}^{\prime }=3$ (or $\lambda
_{ph}=3,\lambda _{ph}^{\prime }=0$), therefore $\lambda =odd$ (only
octupolar transitions allowed);

- transitions between excited vibrational bands are supressed as the change
in the number of vibrational phonons $\Delta n\geqq 2$ (one phonon should be
annihilated and one created for such transitions!).

Note that above selection rules may be expanded if we consider the $K$-mixing
in the nuclear wave functions.

Let's study intra-band transitions in the ground state rotational band when $%
K=K^{\prime }=0$ and the projection $\mu =0$. Obviously $\lambda
_{ph}=\lambda _{ph}^{\prime }=0$ (no phonons in the ground state), therefore
the transition multipolarity $\lambda $ should be even. Additionally, $I$
and $I^{\prime }$ are even numbers ($I,I^{\prime }=0,2,4,...$). From Eq.~%
\eqref{red-mat-even-gen} by doing some simple algebra, and considering that
a factor of 2 arises from the left side of the equation for $\mu=0$, we obtain%
\begin{equation}
\left\langle IK\vert\vert D_{;0}^{\lambda }\vert\vert I^{\prime }K\right\rangle =\sqrt{2I^{\prime }+1}\left\langle I^{\prime
}\lambda K0\vert IK\right\rangle   \label{red-mat-mu0-even}
\end{equation}%
Tamura studied the coupling of states that belong to the ground state
rotational band in even-even nuclei and derived exactly the same formula
(see Eq.~(41) of Ref.~\cite{Tamura:65}). \

\subsection{Odd-$A$ nuclei}

The wave function $\Psi $ for an odd-$A$ nucleus was defined in Eq.~\eqref{wave-odd-gen}, and it reduces to Eq.~\eqref{wave-odd-sym} if we assume axial symmetry (in that case  $K$ is a good quantum number being $K=\Omega$). In the following equations we will omit additional quantum numbers denoted as $\tau $ for simplicity. Using the Eq.~\eqref{wave-odd-gen} for the odd-$A$ nuclear wave function for both the initial and final nuclear states $\langle i\vert$ and $\vert f\rangle$ we can derive the coupled-channel matrix elements in the same way as in the even-even case, using the similar structure of the even-even and odd wave functions.

The Eq.~\eqref{matrix-element} should be slightly modified for odd-$A$ nuclei as follows,
\begin{eqnarray}
\left\langle i\vert V(r,\theta ,\varphi )\vert f\right\rangle  &&={\sum_{K>0}}^{\prime} C_{K}^{I\tau} {\sum_{K^{\prime }>0}}^{\prime} C_{K^{\prime }}^{I^{\prime }\tau^{\prime }}\;\alpha(\nu,\nu^{\prime })\times \notag \\
&&\Bigg\{\sum_{\lambda=0,2,4,..}v_{\lambda }^{(1)}(r)\left\langle IK\vert\vert D_{;0}^{\lambda }\vert\vert I^{\prime}K\right\rangle A(ljI;l^{\prime }j^{\prime }I^{\prime };\lambda J{\frac12})~\delta _{K,K^{\prime }}+ \notag \\
&&v_{2}^{(1)}(r)\times \Bigg\{\left[ \left[ \beta _{2}\right] _{eff}+\left[ \gamma _{20}\right] _{eff}%
\right] ~\left\langle IK\vert\vert D_{;0}^{2}\vert\vert I^{\prime }K\right\rangle A(ljI;l^{\prime }j^{\prime }I^{\prime
};2J{\frac12})~\delta _{K,K^{\prime }}+  \notag \\
&&\left[ \gamma _{22}\right] _{eff}~\left\langle IK\vert\vert \left(
D_{;2}^{2}+D_{;-2}^{2}\right) \vert\vert I^{\prime }\pi ^{\prime }K^{\prime
}\right\rangle A(ljI;l^{\prime }j^{\prime }I^{\prime };2J{\frac12})~+  \label{matrix-element-odd} \\
&&\left[ \beta _{30}\right] _{eff}~\left\langle I\pi K\vert\vert D_{;0}^{3}\vert\vert I^{\prime }\pi
^{\prime }K\right\rangle A(ljI;l^{\prime }j^{\prime }I^{\prime };3J{\frac12})~\delta _{K,K^{\prime }}~+  \notag \\
&&\left[ \beta _{32}\right] _{eff}~\left\langle I\pi K\vert\vert \left(
D_{;2}^{3}+D_{;-2}^{3}\right) \vert\vert I^{\prime }\pi ^{\prime }K^{\prime
}\right\rangle A(ljI;l^{\prime }j^{\prime }I^{\prime };3J%
{\frac12})\Bigg\}\Bigg\}\;,  \notag
\end{eqnarray}%
where the overlapping of the single-particle wave functions is defined by an overlap factor
$\alpha(\nu,\nu^{\prime})=\left\langle \chi_{\nu}\vert \chi_{\nu^{\prime}}\right\rangle$.

Since in practice it is simpler to deal with single-particle states of constant angular momentum $j$, we introduce the expansion \begin{equation}
\chi _{\nu}=\sum_{j\Omega}c_{j\Omega}^{\nu }~\left\vert j\Omega\right\rangle,
\end{equation}
where $\left\vert j\Omega\right\rangle$ are the single-particle basis functions that are eigenfunctions of $j^2$ and $j_{z^{\prime}}$ (spherical basis in the intrinsic $x^{\prime},y^{\prime},z^{\prime}$ system). As pointed out in Ref.~\cite{Hecht62}, in the limit of zero deformation (spherical shell model limit) all but one of the coefficients $c_{j\Omega}^{\nu}$ are equal to zero for a particular state $\nu$, and the summation extends only over the possible values of $j$. The real coefficients $c_{j\Omega}^{\nu }$ determine the mixing of the spherical basis function $\left\vert j\Omega\right\rangle $ into the single-particle wave function $\chi _{\nu }$ in the deformed nuclear field. Those coefficients satisfy the symmetry conditions $c_{j\Omega}^{\nu }=(-1)^{j-1/2}\pi _{\chi}c_{j-\Omega}^{\nu }$ (see p.408 in \cite{Preston75}), where $\pi _{\chi }$ is the parity of the intrinsic wave function.

If we additionally consider
the basis orthogonality condition $\left\langle j\Omega\vert j^{\prime } \Omega^{\prime }\right\rangle =\delta _{jj^{\prime }}\delta _{\Omega \Omega^{\prime }}$ we can write for the overlap factor the following expression:
\begin{equation}
\alpha(\nu,\nu^{\prime})\equiv\sum_{j\Omega}c_{j\Omega}^\nu\;c_{j\Omega}^{\nu^{\prime}}
\label{eq:alpha}
\end{equation}
The expansion coefficients $c_{j\Omega}^\nu$ and $c_{j\Omega}^{\nu^{\prime}}$ are real numbers, and correspond to different single-particle bandheads $\nu$ and $\nu^{\prime}$ that define the inter-band coupling. Note that in the asymmetric case ellipsoidal symmetry imposes additional restrictions on odd-$A$ wave function, in particular $(K-\Omega)$ and $(K^{\prime}-\Omega^{\prime})$ must be even integers. These restrictions are indicated by the apostrophe on the summation symbols ${\sum_{K>0}}^{\prime}$ and ${\sum_{K^{\prime}>0}}^{\prime}$ in Eq.~\eqref{matrix-element-odd}.

Starting from the Eq.~\eqref{red-mat-even-gen}, and considering that $\mu=even$; and $I,$ $I^{\prime }$ and $K$ in odd-$A$ nuclei are half-integer (therefore $\delta _{K,0}=\delta _{K^{\prime },0}=0$, and $I+I^{\prime }=odd$), and replacing the phases $(-1)^{I+\lambda _{ph}}$ by $(-1)^{I-1/2}\pi _{\chi} $ (and $(-1)^{I^{\prime }+\lambda _{ph}^{\prime }}$ by $(-1)^{I^{\prime}-1/2}\pi _{\chi ^{\prime }}$) as discussed in the odd-$A$ nucleus wave function section,
we obtain for the reduced matrix elements for odd-$A$ nuclei the following general expression,
\begin{eqnarray}
\left\langle IK\vert\vert \left[ D_{;\mu }^{\lambda }+(-1)^{\mu }D_{;-\mu }^{\lambda }\right]
\vert\vert I^{\prime }K^{\prime }\right\rangle  &=& \; \sqrt{2I^{\prime }+1} \; \frac{[1+(-1)^{\lambda
}\pi _{\chi }\pi _{\chi ^{\prime }}]}{2} \; \times \notag \\  
&&\big[\left\langle I^{\prime }\lambda K^{\prime }\mu \vert IK\right\rangle +(-1)^{I^{\prime }-1/2}\pi
_{\chi ^{\prime }}\left\langle I^{\prime }\lambda -K^{\prime }\mu \vert IK\right\rangle +
\label{red-mat-odd-gen} \\
&&(-1)^{I-1/2}\pi _{\chi }\left\langle I^{\prime }\lambda K^{\prime }\mu
\vert I-K\right\rangle +(-1)^{I+I^{\prime }-1}\pi _{\chi }\pi _{\chi ^{\prime }}\left\langle I^{\prime
}\lambda -K^{\prime }\mu \vert I-K\right\rangle \big]  \notag
\end{eqnarray}%

For axially-symmetric nuclei $K=\Omega$ become good quantum numbers, and the overlap factor $\alpha(\nu,\nu^{\prime})=\delta_{K,K^{\prime}}$, therefore only multi-band couplings with $\Delta K=0$ are allowed. Axial symmetry is a good approximation for odd-$A$ actinides, where interband coupling strengths are much weaker than in even-even nuclei.

The following selection rules are a consequence of constraints from Eq.~\eqref{red-mat-odd-gen},

- for intraband transitions the initial and final states are the same, i.e., $\chi _{\nu}\equiv\chi _{\nu}^{\prime}$, therefore
$\pi_{\chi }\pi _{\chi ^{\prime }}=+1,$ $\lambda =even$,
and $\alpha(\nu,\nu)=\sum_{j \Omega} (c_{j^\Omega}^\nu)^{2}\equiv 1$ due to the orthogonality condition of the spherical basis;

- for inter-band transitions if $\pi _{\chi }=\pi _{\chi ^{\prime }}$ then $\lambda =even$; if $\pi _{\chi }\neq \pi _{\chi ^{\prime }}$ then $\lambda=odd$. In both cases $\alpha(\nu,\nu^{\prime})\ll 1$. Note that for this case the intrinsic states $\chi _\nu$ and $\chi^{\prime }_\nu$ are different (and the coefficients $c_{j\Omega}^\nu$ and $c_{j\Omega}^{\nu^{\prime}}$ are also different) even if the bandheads have the same {\it approximate} value of $K$.

As we did for the even-even reduced matrix elements, let's study the limiting case when the projection $\mu =0$, and we assume that only
intra-band transitions are possible. As discussed above for intra-band transitions we get $\chi _\nu\equiv\chi ^{\prime }_{\nu^{\prime }}$ and $\pi _{\chi }\pi _{\chi^{\prime }}=+1$, $\alpha(\nu,\nu)\equiv 1$, and $\lambda =even$.
Additionally, $K$ is a positive number ($K=1/2,3/2,...$) for odd-$A$ nuclei, therefore for $K=K^{\prime }$ we obtain that $\left\langle I^{\prime }\lambda -K^{\prime}0\vert IK\right\rangle =\left\langle I^{\prime }\lambda K^{\prime }0\vert I-K\right\rangle \equiv 0$, and we can rewrite the expression \eqref{red-mat-odd-gen} as follows (note that a factor of 2 arises from the left side of the equation for $\mu=0$),
\begin{equation}
\left\langle IK \vert\vert D_{;0}^{\lambda }\vert\vert I^{\prime }K \right\rangle =\frac{\sqrt{2I^{\prime
}+1}}{2}\big[\left\langle I^{\prime }\lambda K0\vert IK\right\rangle +(-1)^{I+I^{\prime }-1}\left\langle I^{\prime
}\lambda -K0\vert I-K\right\rangle \big]  \label{red-mat-mu0-odd}
\end{equation}%
Using the equalities $\left\langle I^{\prime }\lambda -K0\vert I-K\right\rangle =(-1)^{I^{\prime }+\lambda
-I}\left\langle I^{\prime }\lambda K0\vert IK\right\rangle ,$ $(-1)^{2I^{\prime }}=-1$ for $I^{\prime }$
half-integer, and using the fact that $\lambda =even$ we can easily derive

\begin{equation}
\left\langle IK\vert\vert D_{;0}^{\lambda }\vert\vert I^{\prime }K\right\rangle =\sqrt{2I^{\prime }+1}\left\langle I^{\prime
}\lambda K0\vert IK\right\rangle
\end{equation}%
The above derived formula is exactly the same derived by Tamura (see
Eq.~(41) of Ref.~\cite{Tamura:65}) which is valid for ground-state band
transitions both for even-even and odd-$A$ nuclei assuming axial symmetry.

As mentioned above, big differences in the coupling strength are seen for inter-band couplings
between even-even and odd-$A$ cases. These differences reflect the fact that
the inter-band coupling in even-even nuclei is a pure collective transition,
while in odd-$A$ nuclei the corresponding strength is reduced by a
single-particle overlap factor $\alpha(\nu,\nu^{\prime})\ll 1$. This is a consequence of the
very different nuclear structure in even-even and odd-$A$ nuclei; for the
later the couplings between low-lying states occur between rotational bands
built on single-particle excited states, and therefore are much weaker.
It is also worth noting that the overlap factor $\alpha(\nu,\nu^{\prime})$ should be set to
zero for transitions between single-particle bands with bandheads
$K$ and $K^{\prime }$ without overlap (e.g., between the isomeric $K^{\prime
}=1/2^{+}$ and the ground state $K=7/2^{-}$ rotational bands of $^{235}U$).

\newpage
\section*{Appendix 3. Coupling scheme and effective deformations for studied actinides}
{\label{Appendix3} }
The coupled-channel coupling schemes employed in the dispersive optical model calculations for neutron induced reactions on $^{238}$U, $^{232}$Th, $^{239}$Pu, $^{235}$U, and $^{233}$U are tabulated below. For all selected actinides the ground state rotational band was assumed to be described by an axial rigid rotor with static deformations $\beta_{2}$, $\beta_{4}$, and $\beta_{6}$ given in Table~\ref{table:stat-deformations}. The multi-band coupling strength is defined by effective deformations [$\beta_{\lambda\mu}$]$_{eff}$ averaged over vibrational wave functions for the even-even targets $^{238}$U and $^{232}$Th.
For odd-$A$ nuclei, the effective deformations are multiplied by a corresponding single-particle factor $\alpha(\nu,\nu^{\prime})$ reflecting the overlap of the wave functions of single-particle states $\nu$ and $\nu^{\prime}$, with quantum numbers $\simeq K$ and $\simeq K^{\prime}$, respectively. Note that the overlap is different from zero because of the $K$ and $\Omega$-mixing for deformed wave functions characterized by  {\it approximate} quantum numbers $\Omega$ ( $\simeq K$ )  and $\Omega^{\prime}$ ($\simeq K^{\prime}$).

\begin{table}[!tbh]
\caption{The ground-state (GS) deformation parameters of actinides allowing
the best fit of experimental data vs theoretical FRDM deformation parameters
derived by M\"oller and Nix \protect\cite{Moller:95}.}
\label{table:stat-deformations}{\ }
\par
\begin{tabular}[b]{c|c|c|c|c|c|c|c|c|c}
\hline\hline
\T & \multicolumn{3}{c|}{$\beta _{2}$} & \multicolumn{3}{c|}{$\beta _{4}$} & \multicolumn{3}{c}{$\beta _{6}$} \B \\ \hline\hline
  Target & Present & FRDM \cite{Moller:95} & \cite{Capote:08} & Present  & FRDM \cite{Moller:95} & \cite{Capote:08} & Present & FRDM \cite{Moller:95} & \cite{Capote:08} \\ \hline
\T $^{232}$Th & 0.211 & 0.207 & 0.213 & 0.063 & 0.108 & 0.069 &  0.0018 &  0.003 &  0.0017 \\
\T $^{233}$U  & 0.200 & 0.207 & 0.203 & 0.129 & 0.117 & 0.100 & -0.0152 &  0.008 & -0.0300 \\
\T $^{235}$U  & 0.220 & 0.215 & 0.211 & 0.109 & 0.110 & 0.107 & -0.057  & -0.005 & -0.0021 \\
\T $^{238}$U  & 0.230 & 0.215 & 0.228 & 0.062 & 0.093 & 0.062 & -0.0096 & -0.015 & -0.0056 \\
\T $^{239}$Pu & 0.236 & 0.223 & 0.219 & 0.086 & 0.095 & 0.095 & -0.031  & -0.018 & -0.0016 \\ \hline\hline
\end{tabular}%
\end{table}


\begin{table}[tbh]
\caption{Inter-band effective coupling parameters and coupling scheme for $%
n+^{238}$U reaction, 20 coupled levels. The excited vibrational bandhead
energies are 0.680 keV (octupole $K=0^-$), 0.927 keV (quadrupolar $\protect%
\gamma$-band $K=0^+$), 0.993 keV (quadrupolar $\protect\beta$-band $K=0^+$),
and 1.060 keV (non-axial quadrupolar band $K=2^+$). Effective deformations
[$\beta_{\lambda\mu}$]$_{eff}$ are defined after Eq.~\eqref{matrix-element}.
The ground state band and the IAS band static deformations are the same.
IAS need to be coupled for proton induced reactions (the last two listed states).}
\vspace{+2mm}
\label{table:U8-deformations}{\
\begin{tabular}{c|cc|l}
\hline\hline
Excitation energy, MeV & $J^{\pi}$ & $\simeq K$ & Effective deformation \T \\ \hline
0.0000 & $0^+$ & 0 & GS   \\
0.4490 & $2^+$ & 0 & GS   \\
0.1484 & $4^+$ & 0 & GS   \\
0.3074 & $6^+$ & 0 & GS   \\
0.5174 & $8^+$ & 0 & GS   \\
0.6798 & $1^-$ & 0 & $\left[ \beta _{30}\right] _{eff}$=0.062   \\
0.7313 & $3^-$ & 0 & $\left[ \beta _{30}\right] _{eff}$=0.062   \\
0.7759 & $10^+$ & 0 & GS  \\
0.8267 & $5^-$ & 0 & $\left[ \beta _{30}\right] _{eff}$=0.062   \\
0.9270 & $0^+$ & 0 & $\left[ \gamma^2\right] _{eff}$=0.011   \\
0.9663 & $2^+$ & 0 & $\left[ \gamma^2\right] _{eff}$=0.011   \\
0.9664 & $7^-$ & 0 & $\left[ \beta _{30}\right] _{eff}=$0.062 \\
0.9930 & $0^+$ & 0 & $\left[ \beta _2\right] _{eff}$=0.024   \\
1.0373 & $2^+$ & 0 & $\left[ \beta _2\right] _{eff}$=0.024   \\
1.0564 & $4^+$ & 0 & $\left[ \gamma^2\right] _{eff}$=0.011   \\
1.0603 & $2^+$ & 2 & $\left[ \gamma\right] _{eff}$=0.07  \\
1.1057 & $3^+$ & 2 & $\left[ \gamma\right] _{eff}$=0.07  \\
1.1507 & $9^-$ & 0 & $\left[ \beta _{30}\right] _{eff}$=0.062 \\
19.493 & $0^+$ & 0 & IAS \\
19.538 & $2^+$ & 0 & IAS \\ \hline\hline
\end{tabular}%
}
\end{table}

\begin{table}[!tbh]
\caption{Inter-band effective coupling parameters and coupling scheme for $%
n+^{232}$Th reaction, 20 coupled levels. The excited vibrational bandhead
energies are 0.714 keV (octupole $K=0^-$), 0.731 keV (quadrupolar $\protect%
\gamma$-band $K=0^+$), 1.0786 keV (quadrupolar $\protect\beta$-band $K=0^+$%
), and 0.785 keV (non-axial quadrupolar band $K=2^+$).
[$\beta_{\lambda\mu}$]$_{eff}$ are defined after Eq.~\eqref{matrix-element}.
The ground state band and the IAS band static deformations are the same.
IAS need to be coupled for proton induced reactions (the last two listed states).}
\label{table:Th2-deformations}{\
\begin{tabular}{c|cc|l}
\hline\hline
Excitation energy, MeV & $J^{\pi}$ & $\simeq K$ & Effective deformation \T \\ \hline
0.0000 & 0$^+$ & 0 & GS \\
0.4937 & 2$^+$ & 0 & GS \\
0.1621 & 4$^+$ & 0 & GS \\
0.3333 & 6$^+$ & 0 & GS \\
0.5569 & 8$^+$ & 0 & GS \\
0.7144 & 1$^-$ & 0 & $\left[ \beta _{30}\right] _{eff}$=0.047 \\
0.7306 & 0$^+$ & 0 & $\left[ \gamma^2\right] _{eff}$=0.008   \\
0.7742 & 2$^+$ & 0 & $\left[ \gamma^2\right] _{eff}$=0.008   \\
0.7744 & 3$^-$ & 0 & $\left[ \beta _{30}\right] _{eff}$=0.047   \\
0.7853 & 2$^+$ & 2 & $\left[ \gamma\right] _{eff}$=0.008 \\
0.8268 & 10$^+$ & 0 & GS \\
0.8296 & 3$^+$ & 2 & $\left[ \gamma\right] _{eff}$=0.008 \\
0.8730 & 4$^+$ & 0 & $\left[ \gamma^2\right] _{eff}$=0.008 \\
0.8838 & 5$^-$ & 0 & $\left[ \beta _{30}\right] _{eff}$=0.047 \\
1.0429 & 7$^-$ & 0 & $\left[ \beta _{30}\right] _{eff}$=0.047 \\
1.0786 & 0$^+$ & 0 & $\left[ \beta _2\right] _{eff}$=0.01 \\
1.1217 & 2$^+$ & 0 & $\left[ \beta _2\right] _{eff}$=0.01 \\
1.2496 & 9$^-$ & 2 & $\left[ \beta _{30}\right] _{eff}$=0.047 \\
19.4934 & 0$^+$ & 0 & IAS \\
19.5383 & 2$^+$ & 0 & IAS \\ \hline\hline
\end{tabular}
}
\end{table}

\begin{table}[!tbh]
\caption{Inter-band effective coupling parameters and coupling scheme for $%
n+^{239}$Pu reaction, 19 coupled levels. The excited single-particle
bandhead energies are 0.2855 keV ($K=5/2^+$) and 0.4698 keV ($K=1/2^-$).
The excited band at 391.6 keV ($K=7/2^-$) is not coupled as it corresponds to $\Delta K=3$
and parity change, therefore we would need a non-zero $\beta_{33}$ effective deformation, but the octupolar
vibrations with odd-$\mu$ were assumed to be zero.
[$\beta_{\lambda\mu}$]$_{eff}$ are defined after Eq.~\eqref{matrix-element-odd}.
The single-particle overlap factor $\alpha(\nu,\nu^{\prime})$ is given by Eq.~\eqref{eq:alpha}.}
\label{table:Pu9-deformations}{\
\begin{tabular}{c|cc|l}
\hline\hline
Excitation energy, MeV & $J^{\pi}$ & $\simeq K$ & Effective deformation \T \\ \hline
0.00000 & 1/2$^+$   & 1/2 & GS \\
0.00786 & 3/2$^+$   & 1/2 & GS \\
0.05727 & 5/2$^+$   & 1/2 & GS \\
0.07570 & 7/2$^+$   & 1/2 & GS \\
0.16376 & 9/2$^+$   & 1/2 & GS \\
0.19280 & 11/2$^+$  & 1/2 & GS \\
0.28551 & 5/2$^{+}$ & 5/2 & $\alpha(5/2,1/2)\times[\beta_{32}]_{eff}=$0.025 \\
0.31850 & 13/2$^{+}$& 1/2 & GS \\
0.33010 & 7/2$^{+}$ & 5/2 & $\alpha(5/2,1/2)\times[\beta_{32}]_{eff}=$0.025 \\
0.35810 & 15/2$^{+}$& 1/2 & GS \\
0.38740 & 9/2$^{+}$ & 5/2 & $\alpha(5/2,1/2)\times[\beta_{32}]_{eff}=$0.025 \\
0.46200 & 11/2$^{+}$& 5/2 & $\alpha(5/2,1/2)\times[\beta_{32}]_{eff}=$0.025 \\
0.46980 & 1/2$^{-}$ & 1/2 & $\alpha(1/2,1/2)\times[\beta_{30}]_{eff}=$0.062 \\
0.49210 & 3/2$^{-}$ & 1/2 & $\alpha(1/2,1/2)\times[\beta_{30}]_{eff}=$0.062 \\
0.50560 & 5/2$^{-}$ & 1/2 & $\alpha(1/2,1/2)\times[\beta_{30}]_{eff}=$0.062 \\
0.51930 & 17/2$^{+}$& 1/2 & GS \\
0.55620 & 7/2$^{-}$ & 1/2 & $\alpha(1/2,1/2)\times[\beta_{30}]_{eff}=$0.062 \\
0.58300 & 9/2$^{-}$ & 1/2 & $\alpha(1/2,1/2)\times[\beta_{30}]_{eff}=$0.062 \\
0.66110 & 11/2$^{-}$& 1/2 & $\alpha(1/2,1/2)\times[\beta_{30}]_{eff}=$0.062 \\ \hline\hline
\end{tabular}%
}
\vspace{+1cm}
\end{table}

\begin{table}[!tbh]
\caption{Inter-band effective coupling parameters and coupling scheme for $%
n+^{235}$U reaction, 19 coupled levels. The excited single-particle
bandhead energies are 0.3932 keV ($K=3/2^+$), 0.4456 keV ($K=7/2^+$), and
0.6378 keV ($K=3/2^-$).
The excited band built on the isomeric state ($K=1/2^+$) is not coupled to the ground state band
as the isomerism means that the overlap of the corresponding single-particle bandheads is negligible small.
The excited bands at 129.3 keV ($K=5/2^+$) and 332.8 keV ($K=5/2^+$) are not coupled as they correspond to $\Delta K=1$
and parity change, therefore we would need a non-zero $\beta_{31}$ effective deformation, but the octupolar
vibrations with odd-$\mu$ were assumed to be zero.
[$\beta_{\lambda\mu}$]$_{eff}$ are defined after Eq.~\eqref{matrix-element-odd}.
The single-particle overlap factor $\alpha(\nu,\nu^{\prime})$ is given by Eq.~\eqref{eq:alpha}.}
\vspace{+2mm}
\label{table:U5-deformations}
. {\
\begin{tabular}{c|cc|l}
\hline\hline
Excitation energy, MeV & $J^{\pi}$ & $\simeq K$ & Effective deformation \T \\ \hline
0.0000 & 7/2$^-$   & 7/2 & GS \\
0.0462 & 9/2$^-$   & 7/2 & GS \\
0.1030 & 11/2$^-$  & 7/2 & GS \\
0.1715 & 13/2$^-$  & 7/2 & GS \\
0.2491 & 15/2$^-$  & 7/2 & GS \\
0.3385 & 17/2$^-$  & 7/2 & GS \\
0.3932 & 3/2$^{+}$ & 3/2 & $\alpha(3/2,7/2)\times[\beta_{32}]_{eff}=$0.040 \\
0.4267 & 5/2$^{+}$ & 3/2 & $\alpha(3/2,7/2)\times[\beta_{32}]_{eff}=$0.040 \\
0.4394 & 19/2$^{-}$& 7/2 & GS \\
0.4456 & 7/2$^{+}$ & 7/2 & $\alpha(7/2,7/2)\times[\beta_{30}]_{eff}=$0.0455\\
0.4738 & 7/2$^{+}$ & 3/2 & $\alpha(3/2,7/2)\times[\beta_{32}]_{eff}=$0.040 \\
0.5106 & 9/2$^{+}$ & 7/2 & $\alpha(7/2,7/2)\times[\beta_{30}]_{eff}=$0.0455\\
0.5332 & 9/2$^{+}$ & 3/2 & $\alpha(3/2,7/2)\times[\beta_{32}]_{eff}=$0.040 \\
0.5512 & 21/2$^{-}$& 7/2 & GS \\
0.5788 & 11/2$^{+}$& 7/2 & $\alpha(7/2,7/2)\times[\beta_{30}]_{eff}=$0.0455\\
0.6082 & 11/2$^{+}$& 3/2 & $\alpha(3/2,7/2)\times[\beta_{32}]_{eff}=$0.0400\\
0.6378 & 3/2$^{-}$ & 3/2 & $\alpha(3/2,7/2)\times[\gamma_{22}]_{eff}=$0.030\\
0.6645 & 5/2$^{-}$ & 3/2 & $\alpha(3/2,7/2)\times[\gamma_{22}]_{eff}=$0.030\\
0.6902 & 13/2$^{+}$& 7/2 & $\alpha(7/2,7/2)\times[\beta_{30}]_{eff}=$0.0455\\
0.7012 & 7/2$^{-}$ & 3/2 & $\alpha(3/2,7/2)\times[\gamma_{22}]_{eff}=$0.030\\
0.7550 & 9/2$^{-}$ & 3/2 & $\alpha(3/2,7/2)\times[\gamma_{22}]_{eff}=$0.030\\ \hline\hline
\end{tabular}%
}
\end{table}

\begin{table}[!tbh]
\caption{Inter-band effective coupling parameters and coupling scheme for $%
n+^{233}$U reaction, 16 coupled levels. The excited single-particle
bandhead energies are 0.2988 keV ($K=5/2^-$) and 0.3985 keV ($K=1/2^+$).
The excited band at 311.9 keV ($K=3/2^+$) is not coupled as it corresponds to $\Delta K=1$
and no parity change, therefore we would need a non-zero $\beta_{21}$ effective deformation,
but the quadrupole vibrations with odd-$\mu$ were assumed to be zero.
[$\beta_{\lambda\mu}$]$_{eff}$ are defined after Eq.~\eqref{matrix-element-odd}.
The single-particle overlap factor $\alpha(\nu,\nu^{\prime})$ is given by Eq.~\eqref{eq:alpha}.}
\vspace{+2mm}
\label{table:U3-deformations}{\
\begin{tabular}{c|cc|l}
\hline\hline
Excitation energy, MeV & $J^{\pi}$ & $\simeq K$ & Effective deformations \T \\ \hline
0.00000 & 5/2$^{+}$  & 5/2 & GS \\
0.04035 & 7/2$^{+}$  & 5/2 & GS \\
0.09216 & 9/2$^{+}$  & 5/2 & GS \\
0.15523 & 11/2$^{+}$ & 5/2 & GS \\
0.22947 & 13/2$^{+}$ & 5/2 & GS \\
0.29880 & 5/2$^{-}$  & 5/2 & $\alpha(5/2,5/2)\times[\beta_{30}]_{eff}=$0.065 \\
0.31460 & 15/2$^{+}$ & 5/2 & GS \\
0.32080 & 7/2$^{-}$  & 5/2 & $\alpha(5/2,5/2)\times[\beta_{30}]_{eff}=$0.065 \\
0.35380 & 9/2$^{-}$  & 5/2 & $\alpha(5/2,5/2)\times[\beta_{30}]_{eff}=$0.065 \\
0.39760 & 11/2$^{-}$ & 5/2 & $\alpha(5/2,5/2)\times[\beta_{30}]_{eff}=$0.065 \\
0.39850 & 1/2$^{+}$  & 1/2 & $\alpha(1/2,5/2)\times[\beta_{32}]_{eff}=$0.034 \\
0.41170 & 17/2$^{+}$ & 5/2 & GS \\
0.41580 & 3/2$^{+}$  & 1/2 & $\alpha(1/2,5/2)\times[\beta_{32}]_{eff}=$0.034 \\
0.45610 & 5/2$^{+}$  & 1/2 & $\alpha(1/2,5/2)\times[\beta_{32}]_{eff}=$0.034 \\
0.51755 & 19/2$^{+}$ & 5/2 & GS \\
0.63527 & 21/2$^{+}$ & 5/2 & GS \\ \hline\hline
\end{tabular}
}
\vspace{+1cm}
\end{table}


\end{document}